# Direct Derivation of Anisotropic Atomic Displacement Parameters from Molecular Dynamics Simulations in Extended Solids with Substitutional Disorder Using a Neural Network Potential

**Yoyo Hinuma**
Department of Energy and Environment, National Institute of Advanced Industrial Science and Technology (AIST), 1-8-31, Midorigaoka, Ikeda, Osaka 563-8577, Japan
*y.hinuma@aist.go.jp

**ABSTRACT:**
Atomic displacement parameters (ADPs) are crystallographic information that describe the statistical distribution of atoms around an atom site. Anisotropic ADPs by atom were directly derived from classical molecular dynamics (MD) simulations using a universal machine-learned potential. The (co)valences of atom positions were taken over recordings at different time steps in a single MD simulation. The procedure was demonstrated on extended solids, namely rocksalt structure MgO and three thermoelectric materials, $Ag_8SnSe_6$, $Na_2In_2Sn_4$, and $BaCu_{1.14}In_{0.86}P_2$. Unlike the very frequently used lattice dynamics approach, the MD approach can obtain ADPs in crystals with substitutional disorder and explicitly at finite temperature, but not under conditions where atoms migrate in the crystal. The calculated ADP becomes →0 at temperature →0 and the ADP is proportional to the temperature when the atom is in a harmonic potential and the sole contribution to the actual non-zero ADP is from the zero-point motion. The zero-point motion contribution can be estimated from the proportionality constant assuming this Einstein model. ADPs from MD simulations would act as a tool complementing experimental efforts to understand the crystal structure including the distribution of atoms around atom sites.

## 1. Introduction

Atomic displacement parameters (ADPs) are often provided as part of crystallographic structural data and may represent atomic motion, possible static displacive disorder, and thermal vibration. (Trueblood *et al.*, 1996) There was earlier discussion on "thermal vibrations" (Willis & Pryor, 1975), but the International Union of Crystallography now recommends using the term "ADP". (Trueblood *et al.*, 1996) ADPs are generally anisotropic. Calculation of ADPs are important especially to model those of light elements such as H, where obtaining ADPs from X-ray diffraction is difficult, and to provide clues for obtaining a better picture of the actual crystal structure through refinement.

Theoretical derivations of ADPs are typically performed indirectly through lattice dynamics analysis, where the dynamical matrix is obtained and the vibrational, or phonon, frequency of each mode is calculated. Examples of formalisms are found in Refs. (Erba *et al.*, 2013; Madsen *et al.*, 2013; Lane *et al.*, 2012).

The lattice dynamics approach is difficult to apply in systems where disorder of elements on a (sub)lattice plays a critical role. Models with large supercells and/or a very short range order on the (sub)lattice is required and the symmetry is different from the experimental model. The effect of ordering may significantly affect the final results in a small supercell with short range ordering.

Split-site systems are also problematic with lattice dynamics. One example of a split site is a double-well potential with minima at two very close positions. The average position of an atom at this site will be the middle of the two positions at elevated temperatures, while the atom will be at one of the minima at the 0 K limit. Placing the atom at the elevated temperature position leads to an imaginary mode at the Γ point.

In contrast, direct derivation of anisotropic ADPs, assuming a normal (or Gaussian) distribution, from molecular dynamics (MD) simulations is possible even for disordered systems and split site systems. The averages of displacement vector components from the mean atom position, such as $<\Delta\xi^{C}_{i}\Delta\xi^{C}_{j}>$, are simply the (co)variances of atom positions, for instance Cov($\Delta\xi^{C}_{i}+\xi^{C}_{i}$, $\Delta\xi^{C}_{j}+\xi^{C}_{j}$). Here, $\xi^{C}_{i}$ and $\Delta\xi^{C}_{i}$ are the *i*-component of the mean atom position vector and the displacement vector, respectively, in Cartesian coordinates for a given atom. A large number of displacement values for a given crystal can be obtained from MD simulations, and derivation of ADPs from these values is very straightforward. This approach was used in the literature to obtain the anisotropic ADPs of $ND_3$ (Reilly *et al.*, 2007) and benzophenone (Reilly *et al.*, 2013).

The ADP of an individual atom and that of an atom site must be clearly distinguished. The (co)variance of positions for a single atom is obtained in the former, while the (co)variance in the latter is taken over positions of multiple atoms that are each affine transformed such that the images of atoms are at almost the same position. The latter is experimentally observed, while the former is readily obtained using MD.

The direct method can, in principle, derive ADPs for any system. One notable exception is when atoms migrate during the simulations and are not trapped around an equilibrium point. The drawback of the direct method is that each MD simulation gives different results and obtaining very reliable values require a long simulation time and/or a large supercell, which could be computationally expensive using first-principles calculations.

Verifying the accuracy of ADPs is a difficult issue to address. The ADPs in crystal structure refinement absorb systematic errors. Thus, it is highly challenging to obtain accurate ADPs. In X-ray electron density analysis, ADPs obtained separately from X-ray and neutron diffraction data are typically compared to evaluate the accuracy. An extremely careful study on monoclinic $Ni(ND_3)_4(NO_2)_2$ showed average differences of ADPs in non-hydrogen atoms on the order of few 0.0001 $Å^2$, or less than 10%, between 9(1) K XRD and 13(1) K neutron diffraction measurements (Iversen *et al.*, 1996). The ADPs from classical MD simulations suffer some uncertainty from statistical processing in addition to systematic deviations from the true value from the use of approximations and the difference in classical and quantum dynamics.

This paper demonstrates the use of universal machine-learned neural network potential (NNP) MD simulations to obtain the ADP. The calculated ADPs were compared with experimentally reported values. Rocksalt structure MgO and three thermoelectric materials, namely argyrodite structure $Ag_8SnSe_6$ (Takahashi *et al.*, 2024), $Na_2In_2Sn_4$ (Yamada *et al.*, 2023) with a helical tunnel framework structure, and $ThCr_2Si_2$-type phosphide $BaCu_{1.14}In_{0.86}P_2$ (Sarkar *et al.*, 2024b). Some have intrinsic substitutional disorder. The crystal structures of thermoelectric materials are visualized in Fig. 1. The ADPs are especially interesting in the thermoelectric materials because rattling atoms with large ADPs, which may be very anisotropic, could be effective in reducing the thermal conductivity and improving the thermoelectric performance.

## 2. Formalism

Various symbols are used in the literature to describe ADPs, including $\mathbf{U}$, $\mathbf{U}^C$, $\mathbf{B}$, and $\boldsymbol{\beta}$, and these have not always been used consistently. Table 1 summarizes the recommendations from a subcommittee on ADP nomenclature from the International Union of Crystallography (from section 1.3 of Trueblood et al. (Trueblood *et al.*, 1996)). Here, $\mathbf{r}$ is the mean atom position vector and $\mathbf{u}$ is the displacement vector from $\mathbf{r}$. The basis vector lengths of the reciprocal axes are denoted as $a^*$, $b^*$, and $c^*$ or $a^1$, $a^2$, and $a^3$.

The CIF format uses **U**, while the PDB convention adopts $\mathbf{U}^C$.(Grosse-Kunstleve & Adams, 2002) The **U** and $\mathbf{U}^C$ are identical only when the basis vectors are orthogonal to each other. ADPs can be visualized as Oak Ridge Thermal Ellipsoid Plot (ORTEP) ellipsoids, where the semi-axes are along the three eigenvectors of **U** and the lengths are proportional to the corresponding eigenvalue of **U**. (Johnson, 1965) Details of transformation between different ADP definitions are summarized in Appendix A-1.

The following is the proposed procedure to obtain ADPs.

1) Conduct MD simulations.
2) Calculate $\mathbf{U}^C$ for each atom from positions obtained from MD simulations.
3) Average $\mathbf{U}^C$ for atoms related by translational symmetry.
4) If necessary, convert to **U**, change basis vectors of some sites, then average **U** over sites with the same Wyckoff position.
5) Convert to whatever ADP quantity that is convenient.

The ADPs are already standard deviations of numerous atom positions. Providing a standard deviation, often denoted using parenthesis, of a standard deviation (calculated ADP) from a single simulation is mathematically not sound.

Suppose there are 1 million data points from a single MD simulation. The data points can be split into 10 sets with 100 K data points, and the standard deviation of **U** obtained from 10 sets of 100 K data points can be obtained. This standard deviation would be different from 100 sets of 10 K data points. The standard deviation depends on how the data points are split, thus it is not a well-defined value. However, the **U** of the 1 M data points are the same as both the average **U** of 10 sets of 100 K data points and the average **U** of 100 sets of 10 K data points, thus the **U** obtained from all data points is a well-defined value.

The averages and standard deviations of **U** over different runs with different atom orderings, and those of $\mathbf{U}/T$ over different temperatures $T$ are meaningful.

## 3. Methodology

### 3.1 MD procedure

MD simulations were conducted using the commercially available Matlantis package from Preferred Networks with their universal PreFerred Potential (PFP) (Takamoto *et al.*, 2022) version 7.0.0, a NNP trained on the Perdew-Burke-Ernzerhof (PBE) generalized gradient approximation (GGA) to density functional theory (DFT) (Perdew *et al.*, 1996). Diverse compounds consisting of any of all 96 elements lighter than Cm inclusive may be modeled using the same potential.

Adopting a potential model instead of first-principles calculations results in accumulation of atom positions with a pace that is orders of magnitude faster. Too much computational resources are necessary when using first-principles calculations to obtain a sufficient amount of atom position information for derivation of ADPs with reasonable precision. On the other hand, classical force fields or potentials are only available for a limited number of systems, and fine-tuning classical force fields or potentials for each crystal in consideration requires a high level of expertise. Additionally, much time is necessary for fitting and verification. Using an off-the-shelf universal potential applicable to a wide variety of systems is a very practical approach, at least as a first attempt to newly explore a crystal.

Various universal machine learning NNPs were proposed in recent years (Jacobs *et al.*, 2025). The PFP with the Matlantis package was chosen partly because it has a long history with continuous updates by dedicated researchers. The developers trained the potential on a proprietary data set exceeding 60 million configurations using the VASP code (Kresse & Furthmüller, 1996; Kresse & Joubert, 1999). It is not overfitted to a particular published data set such as the Materials Project. The PFP is available as a “take it or leave it” potential; the user cannot modify it.

The canonical, or constant number of atoms, volume, and temperature (NVT), ensemble was used with a Nosé–Hoover thermostat (Nosé, 1984; Hoover, 1985). The objective of this paper is to demonstrate direct derivation of ADPs from MD simulations using a reasonable energy and force calculator, and fine-tuning the NNP for individual compounds is outside the scope.

Structural and calculation details of the considered crystals are described below.

### 3.2 MgO simulations

The experimental ADPs of MgO were compared with theoretical values obtained using the proposed procedure. MgO takes the rocksalt structure with space group type $Fm\overline{3}m$ (number 225). The Mg and O occupy one Wyckoff position each (4*a* and 4*b*), both without variable parameters. The ADPs are isotropic.

The MgO conventional unit cell contains four Mg and four O atoms, respectively. The computational lattice parameter at 0 K is 4.2567 Å, which was used for MD simulations. Supercells of the conventional cell with sizes 3×3×3, 4×4×4, 5×5×5, 6×6×6, and 7×7×7 were obtained and used.

The effect of the lattice expansion on the ADPs was also studied. The lattice parameter at 300 K is 0.15% larger than at 20-100 K, according to the experimental volume per mol data in Ref. (White & Anderson, 1966). Therefore, additional calculations were conducted on 5×5×5 supercells where the lattice parameter was increased by 0.15%.

MD simulations were conducted with a time step of 2 fs, and the number of steps was 50,000 steps (100 ps). Positions were recorded every 100 steps, but the positions for the first 10,000 steps (20 ps, 100 position recordings, hereafter equilibration steps) were discarded to account for initial shifting of atoms to attain equilibration. The temperature $T$ was varied between 50 to 500 K in 50 K intervals.

### 3.2. $Ag_8SnSe_6$ simulations

The unit cell of $Ag_8SnSe_6$ contains 30 atoms and its space group type is $Pmn2_1$ (number 31). The experimental lattice parameters at 300 K are $a$ = 7.91440 Å, $b$ = 7.82954 Å, and $c$ = 11.05912 Å, which were used for MD simulations. The Ag atoms rattle and are of interest in this study. There are five symmetrically different Ag sites. The Wyckoff positions of sites Ag1, Ag2, and Ag3 are 4*b*, and those of Ag4 and Ag5 are 2*a* (Takahashi *et al.*, 2024).

2×2×2 and 3×3×3 supercells were built for MD simulations of $Ag_8SnSe_6$. The time step was 2 fs, and the number of steps was 155,000 and 60,000 steps, respectively (310 and 120 ps, respectively). Positions were recorded every 100 steps, but the positions for the first 20,000 steps (40 ps, 200 position recordings) were discarded as equilibration steps. The number of Na position data is the same in the two calculations (16×2×2×2 Ag atoms × 1350 position recordings and 16×3×3×3 Ag atoms × 400 position recordings). The temperature $T$ was varied between 50 to 300 K in 50 K intervals.

The **U** for each Ag atom in a MD simulation was obtained by taking the (co)variances of atom positions from position recordings. The symmetrically equivalent coordinate triplets for 4*a* sites are (*x*,*y*,*z*), (-*x*+1/2,-*y*,*z*+1/2), (*x*+1/2,-*y*,-*z*+1/2), and (-*x*,*y*,*z*). The **U** for the second, third, and fourth types of atoms can be transformed to **U** for the first type, which is reported in this paper, using a matrix **R** (equation 16) that is a diagonal matrix with diagonal components (-1,-1,1), (1,-1,1), and (-1,1,1), respectively. For 2*a* sites with equivalent coordinate triplets (0,*y*,*z*) and (0,-*y*,*z*+1/2), the diagonal matrix **R** of the latter has diagonal components (1,-1,1). The final **U** value of each Wyckoff position of Ag was derived by averaging the **U**, for each atom, over all atoms in all simulations.

### 3.3. $Na_2In_2Sn_4$ simulations

The unit cell of $Na_2In_2Sn_4$ contains 16 atoms and its space group type is $P2_12_12_1$ (number 19). The experimental lattice parameters at 300 K are $a$ = 6.3091 Å, $b$ = 6.5632 Å, and $c$ = 11.3917 Å, which were used for MD simulations. There are four 4*a* sites, where one is fully occupied by Na and the other three are shared by In and Sn with a 1:2 ratio. Na has high anisotropy and rattle in this compound. (Yamada *et al.*, 2023)

For $Na_2In_2Sn_4$, 4×4×2 supercells were built containing 128 Na sites and 384 In/Sn sites. 10 supercells, where 128 In and 256 Sn atoms were randomly assigned to the 384 In/Sn sites, were built as initial structures. The time step was 2 fs, positions were recorded every 100 steps, and 55,000 steps (110 ps, 500 position recordings) were considered. The first 5,000 steps (10 ps, 50 position recordings) were discarded as equilibration steps. The temperature $T$ was varied between 50 to 300 K in 50 K intervals.

The final **U** was obtained similarly as in $Ag_8SnSe_6$. The symmetrically equivalent coordinate triplets for 4*a* sites are (*x*,*y*,*z*), (-*x*+1/2,-*y*,*z*+1/2), (-*x*,*y*+1/2,-*z*+1/2), and (*x*+1/2,-*y*+1/2,-*z*). The **U** for the second, third, and fourth types of atoms can be transformed to **U** for the first type, which is reported in this paper, using a matrix **R** (equation 16) that is a diagonal matrix with diagonal components (-1,-1,1), (-1,1,-1), and (1,-1,-1), respectively.

### 3.4. $BaCu_{1.14}In_{0.86}P_2$ simulations

The unit cell of $BaCu_{1.14}In_{0.86}P_2$ contains 10 atoms and its space group type is $I4/mmm$ (number 139). The experimental lattice parameters at 175 K are $a = b$ = 4.0073 Å and $c$ = 13.451 Å, which were used for MD calculations. Cu and In share a 4*d* site and P occupies a 4*e* site. Ba mainly resides at the 2*a* site, but the Ba site is reported as triple-split; 82% of Ba is at the 2*a* site, while 18% of Ba are at a 4*e* site very close to the 2*a* site (at $z = \pm 0.02$ compared to $z = 0$ at the 2*a* site).

6×6×2 supercells were used for MD simulations, which contain 144, 164, 124, and 288 Ba, Cu, In, and P atoms, respectively, or 72($BaCu_{1.139}In_{0.861}P_2$). 10 supercells were prepared as initial structures. The Cu and In atoms were randomly assigned to the 288 Cu/In sites. All Ba were initially positioned at the center of the triple split sites, namely the 2*a* site, assuming that Ba can move to the 4*e* site, as necessary, during the initial equilibration steps. The time step was 2 fs, positions were recorded every 100 steps, and 55,000 steps (110 ps, 500 position recordings) were considered. The first 5,000 steps (10 ps, 50 position recordings) were discarded as equilibration steps. The temperature *T* was varied between 25 to 300 K in 25 K intervals.

The final **U** was obtained similarly as in $Ag_8SnSe_6$. The *I*4/*mmm* symmetry forces $U^{11}=U^{22}$ and $U^{23}=U^{13}=U^{12}=0$, although this is not exactly attained with statistical handling of atom positions. The quantities $(U^{11}+U^{22})/2$ (simply denoted as $U^{11}$ for brevity), $U^{33}$, and the isotropic $U_{\mathrm{iso}}= (U^{11}+U^{22}+U^{33})/3$ were evaluated in this study.

### 3.5 Similarity of ADPs

An obvious way to discuss the similarities of isotropic ADPs is simply comparing the *U* values. However, comparing anisotropic **U** with different principal axis directions is not straightforward, especially in degenerate cases where the principal axes can be taken differently (an extreme case is isotropic **U**).

According to Ref. (Whitten & Spackman, 2006), a measure of the overlap between probability density functions from two ADPs, $\mathbf{U}_1$ and $\mathbf{U}_2$, is

$$R_{12} = \frac{2^{3/2}\left(\det \mathbf{U}_1^{-1}\mathbf{U}_2^{-1}\right)^{1/4}}{\left[\det\left(\mathbf{U}_1^{-1}+\mathbf{U}_2^{-1}\right)\right]^{1/2}} \qquad (1)$$

and the similarity index is defined as

$$S_{12} = 100\left(1-R_{12}\right). \qquad (2)$$

For isotropic **U** with $U_{\mathrm{iso1}}$ and $U_{\mathrm{iso2}}$,

$$R_{12} = \left(\frac{2}{U_{\mathrm{iso1}}+U_{\mathrm{iso2}}}\sqrt{U_{\mathrm{iso1}}U_{\mathrm{iso2}}}\right)^{3/2}, \qquad (3)$$

which is the ratio of the geometric to arithmetic mean raised to the power of 3/2.

## 4. Results and discussion

The computational ADPs are those of individual atoms (ADP by atom) unless noted otherwise, while the experimental ADPs are as-reported values of ADP by site.

### 4.1. MgO

Fig. 2 shows the $U_{iso}$ versus $T$ for Mg and O (the subscript $_{iso}$ is dropped for brevity hereon in this section). The lines are linear regressions of $U \propto T$, and this trend is found over all temperature ranges and for all supercells.

The convergence of $U_{iso}$ with respect to sampling time may be judged by plotting the $U_{iso}$ over certain time periods in a long MD run. There are 500 position recordings from the 50,000 step (100 ps) MD simulations, and the 10 $U_{iso}$ obtained from the 50($n$-1)+1 -th to 50$n$ -th position recordings are plotted for $1 \le n \le 10$ in Fig. 3. The $U_{iso}$ is converged when the $U_{iso}$ is roughly the same value over different $n$. The $U_{iso}$ for $n$=1 is clearly larger than those from other $n$, and the $n$=1 and 2 results are discarded as "still in equilibration and not converged with respect to sampling time.

Fig. 4 plots $U/T$ versus $T$. Ideally, the proportionality factor should be the same over the entire temperature and over all supercells with the same lattice parameters. $U/T$ for $T \le 150$ K calculations tend to be larger than for $T \ge 200$ K calculations in supercells other than 3×3×3, thus the $U/T$ averaged over $T \ge 200$ K points are considered afterwards. There is a 11% difference between $U/T$ of 3×3×3 and 7×7×7 supercells for both Mg and O, while this difference decreases to 2% between 5×5×5 and 7×7×7 supercells for both Mg and O. Therefore, the 5×5×5 supercell is reasonably convergence with regard to size.

The standard deviation of $U/T$ normalized by the average $U/T$ over the seven points between 200 and 500 K is discussed next. With the exception of 5% for Mg and O in the 5×5×5 supercell, the value is less than 3% for both Mg and O in the other four supercell sizes. Increasing the lattice parameter of the 5×5×5 supercell by 0.15% results in a ~1% increase in the $U/T$, thus considering thermal expansion does not result in a substantial difference in $U$, especially near room temperature.

Experimental determination of ADPs of both Mg and O is possible using XRD partly because their atomic numbers, $Z$, are close to each other ($Z$=12 and 8 for Mg and O, respectively). The reported values of $U$ from XRD or electron diffraction at room temperature are 0.0038-0.0040 and 0.0042-0.0046 $Å^2$ for Mg and O respectively. (Lawrence, 1973; Sasaki *et al.*, 1979; Tsirelson *et al.*, 1998)

The fitted values of 0.0037 and 0.0034 $Å^2$ for Mg and O, respectively, at 298 K with the 5×5×5 supercell (the same value is obtained with or without lattice parameter expansion of 0.15%), slightly underestimate experimental values. However, applying a correction accounting for zero-point motion based on the Einstein model increases $U$ to 0.0042 $Å^2$ for both Mg and O, with or without lattice parameter expansion (the correction is discussed detail in section 4.5). This corrected value for O is much closer to the experimental results.

### 4.2. $Ag_8InSe_6$

Figs. 5(a,b) show the ADPs for Ag obtained from 2×2×2 and 3×3×3 supercells, respectively. The isotropic ADP ($U_{iso}$) is given for Ag2, Ag3, and Ag5 while the three eigenvalues of the ADP are provided, as $U_1<U_2<U_3$, to be consistent with Fig. 3. in Takahashi et al. (Takahashi *et al.*, 2024), which is reproduced as Fig. 5(c) with the same symbols and scales as in Figs. 5(a,b). The 2×2×2 and 3×3×3 supercell results are very similar except for 300 K, suggesting good convergence, and have roughly the same values as the experimental results in Fig. 5(c), However, there are minor differences. Ag3 $U_1$ and $U_2$ approaches 0 as temperature $T \rightarrow 0$ in calculations, but this is not the case in experiment.

The calculated 2×2×2 supercell, 3×3×3 supercell, and experimental results in Fig. 5 show qualitatively similar trends, but the exact values can differ than more than 0.01 $Å^2$, which could be considered a very large value. Computational values can be refined by adding more data points by increasing the simulation time. The ADPs from the two supercells in Fig. 5 are different by less than 10 % except for Ag3 $U_3$ at 300 K, and the systematic difference between calculations and experiments cannot be addressed by calculating more data points. Developing and/or using an energy and forces calculator other than PFP may result in a better agreement, but this is outside the scope of this study.

The principal semi-axis lengths of the ellipsoid, which are the three eigenvalues of the matrix **U**, help understand the shape of the ellipsoid. The calculated $U$ values are $U_1<U_2<<U_3$ and $U_1<<U_2<U_3$ in Ag1 and Ag3, respectively, resulting in a cigar-like (prolate) and saucer-like (oblate) ellipsoids, respectively, as expected from experimental results. The computational Ag ADPs at 300 K based on the 3×3×3 supercell is shown in Fig. 1(b). The shapes of the Ag1 and Ag3 ellipsoids are consistent with the experimentally obtained ellipsoids in Fig. 1(a). The calculated Ag1 $U_1$, Ag1 $U_2$, Ag3 $U_2$, and Ag5 $U_{iso}$ are proportional to temperature over the entire shown temperature range, as is indicated in the linear fit that passes through the origin in Figs. 5(a,b). Other calculated **U** values appear to approach $\mathbf{U}\rightarrow 0$ in the limit $T\rightarrow 0$ with the clear exception of Ag3 $U_3$. The Ag3 $U_3$ value for 50 K is surprisingly larger than the 100 K value in the 2×2×2 supercell.

The Ag1 and Ag3 sites are almost three-fold trigonal planar coordinated. The experimental Ag-Se distances for Ag1 are 2.653, 2.659, and 2.707 Å, while those for Ag3 are 2.541, 2.687, and 2.774 Å, respectively. The chemical pressure from the very short Ag3-Se distance of 2.541 Å strongly motivates Ag to rattle in the direction out of the coordination plane (Takahashi *et al.*, 2024), which can effectively result in a split site. This rattling mechanism caused by from "retreat from stress" is found in tetrahedrites and tennantites $(Cu,Zn)_{12}(Sb,As)_4S_{13}$. (Suekuni *et al.*, 2018) The chemical pressure is much weaker for Ag1, thus Ag1 can be a non-split site while a similarly coordinated Ag3 may be a split site in the direction almost normal to the three-fold coordination plane. Experimentally, all of $U_1$, $U_2$, and $U_3$ of Ag3 do not approach $\mathbf{U}\rightarrow 0$ in the limit $T\rightarrow 0$, which is also a hint of site splitting.

Fig. 6 is a schematic of a double-split site for qualitative discussion on the temperature ($T$) dependence of $U$ by atom. The energy is proportional to $T$. At very low $T$ ($T_1$), the atom is trapped in one of the wells, resulting in a small $U$. The $U$ is very large at $T$ ($T_2$) that allows atoms to move between the wells, for example by thermal fluctuation or tunneling, but is sufficiently low that atoms still reside in one of the wells. This is because the atom is typically located far from the average position between the wells. Further increasing $T$ such that the atom is effectively in a single well ($T_3$) results in a substantial decrease in $U$ because the atom can now be at the center of the well. Gradually increasing $T$ ($T_4$) results in a gradually increasing $U$. The $U$ by atom and $U$ by site should be almost the same for $T \geq T_2$, but, at $T_1$, the $U$ by site is expected to be much larger than $U$ by atom because atoms can occupy both wells of the site.

Tables 2 and 3 show the elements of **U**, the eigenvalues $U_1$, $U_2$, and $U_3$, $U_3/U_1$, and $U_{iso}$ for all Ag sites at 200 K and 300 K, respectively, from the 3×3×3 supercell simulations. Ref. (Takahashi *et al.*, 2024) used anisotropic **U** for Ag1 and Ag3 because this dramatically improved the $R_{wp}$ of experimental data at 300 K. Calculations can provide anisotropic **U** for all Ag simultaneously and independently without the need for repeated refinement attempts. The calculated value of a measure of anisotropy, $U_3/U_1$, at 300K is roughly 4 for Ag1, Ag3, and Ag5 (Table 3). The $U_{iso}$ of Ag3 is roughly double of Ag1 and Ag5, while the number of Ag5 atoms is half of Ag1. Therefore, considering anisotropy of Ag5 might result in a smaller improvement of $R_{wp}$ compared to Ag1 and Ag 3. Ag2 and Ag4 are less anisotropic than Ag1, Ag3, and Ag5 because of their smaller $U_3/U_1$, thus using anisotropic **U** would not improve $R_{wp}$ significantly. The trends are similar for both 200 K (Table 2) and 300 K (Table 3).

The convergence of **U** in the 3×3×3 supercell simulation was checked by comparing the similarity index between **U** from position recordings 201 to 400 and 401 to 600 (the initial 200 are discarded). The similarity index was 0.02, 0.07, 0.06, 0.04, and 0.16 for Ag1 to Ag5, respectively. This is about one order of magnitude smaller than the similarity index for Ag1 and Ag3 between the experimental and calculated values (position recordings between 201 and 600), which are 0.67 and 0.43, respectively. Therefore, the **U** values are regarded as converged with respect to sampling time.

### 4.3. $Na_2In_2Sn_4$

Fig. 7 shows the eigenvalues of experimental(Yamada *et al.*, 2023) and calculated **U**, denoted as $U_1<U_2<U_3$, for Na and In/Sn1 sites. ($U_{aniso_a}$ is used instead of $U_3$ in the original reference.) The results for In/Sn2 and In/Sn3 sites are very similar to the In/Sn1 sites and are not shown. The In and Sn **U** are calculated separately, although one **U** for In and Sn combined is obtained experimentally.

The experimental and computational **U** values of Na are comparable to each other, and $U_3$ of Na is significantly larger than $U_1$ and $U_2$ for all temperatures, implying an almost cigar-shaped spheroid. (Fig. 7(a)) The experimental $U_1$, $U_2$, and $U_3$ are roughly proportional to $T$ (linear regressions passing through the origin are shown in black lines). The calculated **U** is roughly the same as the experimental **U**, but the details are slightly different. All of $U_1$, $U_2$, and $U_3$ are almost proportional to $T$ up to about 150 K, but consistently becomes larger than the linear regression of 50, 100, and 150 K values that passes through the origin (shown as red solid lines below 175 K, extrapolations to higher temperature shown with dashed lines above 175 K).

This deviation from proportionality in calculations is even more profound in In/Sn1 sites. The calculated $U_3$ of In1 is comparable to the $U_3$ of Na at 300 K, which is very different from the experimental results! (Fig. 7(b)) However, the experimental and computational **U** values are relatively close to each other at 50 and 100 K (Fig. 7(c), which is an enlargement of Fig. 7(b) at low **U**).

A large **U** value results in a large displacement from the equilibrium position. The standard deviation of the atom position, σ, is simply the square root of **U**. A $U$ of 0.09 $Å^2$ along a certain direction corresponds to a σ of 0.3 Å. Assuming a normal distribution, the atom is more than 3σ = 0.9 Å away from the average position for 0.3% of the time. This 0.9 Å is roughly one-third of the In/Sn-In/Sn bond length (~2.87 Å).

In the author's previous study on $LaH_{2.75}O_{0.125}$ (Hinuma, 2025), the isotropic ADP of O, $U_{iso}$ (denoted as $<\Delta r^2>$ in the reference), is roughly proportional to temperature below ~0.02 $Å^2$ but becomes much larger than what is expected from the proportionality trend above this threshold $U_{iso}$ value. The $U_{iso}$ of La is proportional up to ~0.03 $Å^2$, which covers the entire considered range of $T$. This $LaH_{2.75}O_{0.125}$ is known as a very good H ion conductor. The H and O, together with vacancies, share the same sublattice in $LaH_{2.75}O_{0.125}$, thus O may easily move away from the original site after a very small displacement on the order of ~0.1 Å from the equilibrium position.

The mean square displacement (MSD) of an atom trapped near the equilibrium point becomes a constant regardless of time. This MSD is the ADP when the initial position of the atom is the equilibrium point. However, the mean square displacement is proportional to time in a diffusing atom under Brownian motion, and the proportionality factor is two times the dimension times the diffusion coefficient. For very slowly moving atoms, the moving of some atoms away from the original equilibrium position can be detected but the diffusion coefficient cannot be derived with reasonable precision with a realistic simulation time. MD simulations find that In/Sn atoms can move away from the equilibrium point at above 175 K, based on MSD increasing with time and/or too large **U** values, and thereby the proposed algorithm concedes that it is not applicable in this material. Assigning atoms to the nearest site might be possible, for example by Gaussian mixture modeling or Voronoi tessellation, and atom sites must be correctly identified or provided explicitly.

Inspection of atom movements during MD simulations of $Na_2In_2Sn_4$ revealed significant displacements of atoms at $T \geq 200$ K. The increase in **U** above the proportionality trend in calculations, but not in experiment, is caused by the difference in how **U** is obtained. Experimentally, the displacement of atoms is the distance to the nearest atom site, and the same diffusing atom may be assigned to different sites as time progresses. In contrast, the displacement in calculations is always the distance to the original atom site. The calculated **U** overestimates the actual **U** when atoms can move away from the original site.

For the record, Table 4 shows calculated **U** values of Na together with reported experimental values at 200, 250, and 300 K.(Yamada *et al.*, 2023) The calculated and experimental values for 200, 250, and 300 K are consistent with each other, although the sign is different in some off-diagonal **U** coefficients.

### 4.4. $BaCu_{1.14}Cu_{0.86}P_2$

Table 5 shows the calculated **U** of $BaCu_{1.14}In_{0.86}P_2$ at 175 K together with experimental results (Sarkar *et al.*, 2024b) at 173 K. In the reference, the same $U_{iso}$ was provided for each split Ba site, and anisotropic **U** was not given for Ba. Anisotropic **U** was provided for the Cu/In and P sites. Experimentally, the Cu/In site has the largest **U** and is moderately anisotropic with $U^{33}/U^{11}$=1.38, while the **U** of P is very anisotropic with $U^{33}/U^{11}$=2.24 and is slightly smaller than that of Cu/In. The calculations underestimate experimentally determined **U**. Notably, the calculated $U^{33}$ of Cu/In and P are roughly one-half and one-third of the experimental $U^{33}$, respectively.

Fig. 8 shows the calculated anisotropic **U** of Ba and Cu and isotropic $U_{iso}$ of In and P. The $U_{iso}$ of In and P are almost the same value for all temperatures $T$. All atoms are in an almost harmonic potential, with the linear regressions passing very close to the origin although not required to do so. There were no migrating atoms. The ± 1 standard deviation over the 10 MD runs are shown. The standard deviation of Cu is large because of the large $U$ value and relatively large (standard deviation) / (mean average) ratio.

Assuming $U \propto T$ over the temperature range, the (standard deviation) / (mean average) over all considered temperatures was 4.6% or less for all of $U^{11}$, $U^{33}$, and $U_{iso}$ of all elements (the mean averages and standard deviations are given in Table 6). The correction to **U** from zero-point motion based on the Einstein model (details in Appendix **A-2,** $U_{lowT}$ and $T_C$ in equations 29 and 30, respectively) was calculated. The correction is at most 11% for P and 4% within other elements, respectively. The corrected **U** is given in Table 5.

The difference in **U** between experiment and calculations in Table 5 arises from how the values were derived. The experimental ADPs are by site, while the calculated ADPs in Table 5 and Fig. 8 are the averages of ADPs by atom. Therefore, the $U^{33}$ by site was additionally derived using a histogram of $z$-coordinates. Atoms with coordinates outside of $-0.1<z<0.4$ were translated to this range in integer multiples of 0.5; note that $BaCu_{1.14}In_{0.86}P_2$ is a body-centered crystal.

Figs. 9(a-d) show the histograms for Ba, Cu, In, and P, respectively, for $T$ = 50, 175, and 300 K. The bin size of the $z$-coordinate is 0.001. There are 144 atoms × 500 position recordings × 10 supercells = 720,000 total positions for Ba. Only the $z \approx 0.14$ peak is shown for P (there is another peak at $z \approx -0.14$, which is a mirror image around $z = 0$, that is not shown). The normal distribution regressions and their standard deviations are also given, which can be used to obtain the ADP by site. The histogram for Cu in Fig. 9(b) cannot be described well by a single normal distribution.

The curve for Ba has a single peak, and the experimentally suggested triple well scenario with small peaks at $\Delta z = \pm 0.02$ (Sarkar *et al.*, 2024b) is clearly denied. In contrast, Cu, but not In, shows a triple peak with additional peaks at $\Delta z \approx \pm 0.025$. Figs. 9(e, f) show the histograms of Cu and Cu/In combined, respectively, for $T$ = 50 and 175 K and the regressions

$$y = A\left[p\exp\left\{-\frac{(z-0.25)^2}{2\sigma^2}\right\} + \frac{1-p}{2}\exp\left\{-\frac{(z-0.25-\Delta z)^2}{2\sigma^2}\right\} + \frac{1-p}{2}\exp\left\{-\frac{(z-0.25+\Delta z)^2}{2\sigma^2}\right\}\right]. \quad (4)$$

The values to be fitted are a scaling constant $A$, the occupancy ratio of the main peak $p$, the standard deviation of the peaks σ, and the position of the additional peaks Δz. Equation 4 provides a very good fit, and the parameters used in fitting are given in Table 7. Very interestingly, the $p$ for Cu/In at 175 K is 0.811, which is the experimental occupancy of the main peak of Ba. The Δz of Ba in the experiment is 0.021, while the Δz of Cu/In in the calculation is a very similar value of 0.026.

Fig. 10 plots the calculated $U^{33}$ ADP by atom and by site. The site ADP is obtained as $U^{33}_{\text{site}} = c^2\sigma^2$ where $c$ is the lattice parameter and σ is the standard deviation of the histogram. The ADP was obtained for Cu only, In only, and Cu/In combined for the 4$d$ site. A single normal distribution and a superimposition of three normal distributions with same σ as in equation 4, which models a triple-split site, were calculated with Cu and Cu/In sites.

In Fig. 10, all shown $U^{33}$ are well described with linear fits. There is no contribution to $U$ except for the zero-point motion that is not reflected in the calculations. The non-zero $U^{33}$ by site but zero $U^{33}$ by atom at $T \rightarrow 0$ implies a significant contribution to $U$ by site from disorder of Cu/In sites, which causes atoms to move away from the average atom sites defined as points. This is in addition to any contributions from zero-point motion that is not reflected in the calculations.

The single crystal XRD data in Ref. (Sarkar *et al.*, 2024a) (CCDC Number: 2285845) was re-refined using a model suggested by ADP calculations, which has a triple-split Cu site and no splitting in Ba, In, and P sites. The results are summarized in Table 8. The $R_1$ [$I$>2σ($I$)] and $R_1$ [all data] of the re-refinement are 0.0298 and 0.0318, which are better than 0.031 and 0.033, respectively, in the original reference (Sarkar *et al.*, 2024b). However, the ADPs in the re-refinement are not consistent with the calculated values in Table 5.

### 4.5. Behavior of *U* at *T*→0

The Einstein model is considered first, where atoms are each isolated in a harmonic potential and the atoms do not explicitly interact with each other. The relation $U \propto T$ holds between an ADP component $U$ and temperature $T$ at sufficiently high temperatures. On the other hand, $U$ converges to a non-zero value, $U_{\mathrm{low}T}$, at very low temperatures where the zero-point motion cannot be ignored. A crossover temperature $T_c$ is defined, which is the $T$ corresponding to $U_{\mathrm{low}T}$ assuming that the $U \propto T$ relation holds down to 0 K. This $T_c$ can be obtained once a combination of $U$ and $T$ in the $U \propto T$ regime is obtained. (See Appendix A-2 for mathematical details for this section). Using the $U/T$ proportionality factor is preferred, if available, instead of a single combination.

Experimental results of $U_3$ for In in $Na_2In_2Sn_4$ are proportional to temperature down to 90 K (see Fig. 4(b) in (Yamada *et al.*, 2023)). Using values of $U_3$ = 0.15 Å$^2$ at 250 K in the figure, $U_{\mathrm{low}T}$ = 0.008 Å$^2$ and $T_c$ = 13 K. Therefore, effects from the zero-point motion is not relevant in the temperature range studied in this system.

The heat capacity $C$ in the Einstein model is not proportional to $T^3$ at $T \rightarrow 0$, although experimentally $C \propto T^3$ is often found. In contrast, the Debye model gives $C \propto T^3$ at $T \rightarrow 0$. In the Debye model, the $U$ at $T \rightarrow 0$ converges to a non-zero $U$ value, $U_{\mathrm{low}T_\mathrm{D}}$, and $U \propto T$ at sufficiently high temperature, as in the Einstein model. The crossover temperature $T_{c_D}$ can be defined similarly, which is 1/4 of the Debye temperature in the simplest approximation. The $U_{\mathrm{low}T_\mathrm{D}}$ and $T_{\mathrm{c_D}}$ are $\sqrt{3}/2 \approx 87\%$ smaller than $U_{\mathrm{low}T}$ and $T_\mathrm{c}$, respectively.

The Einstein and Debye models are totally different assumptions on atom vibrations in the crystal. Different atoms can have different characteristic frequencies in different directions in the Einstein model, while there is only one Debye temperature in a crystal.

An elastic neutron scattering study reports the Debye temperature of MgO as 743 ± 8 K.(Beg, 1976), corresponding to $T_{\mathrm{c_D}}$ = 186 K. The corresponding Einstein model crossover temperature, $T_\mathrm{c}$, would be roughly 214 K. The $U$ based on the Einstein model and the $U \propto T$ proportionality factor of the 5×5×5 supercell model are shown as green dashed lines in Fig. 2. The $T_\mathrm{c}$ for Mg and O are 187 and 244 K, respectively, which are close to the 214 K inferred from the experimental Debye temperature.

Looking from the other side, the NNP MD simulations in this study can clearly identify whether the experimental non-zero $U$ at $T \rightarrow 0$ is solely the consequence of zero-point motion in an Einstein model or are there contributions from something else, such as configurational disorder or a split site. Therefore, ADPs estimated from NNP MD acts as a probe that complement experimental observations.

Why does the $U \propto T$ proportionality persist at very low $T$ in the NNP MD simulations, but not in experiment? This paradox can be resolved easily. The NNP MD is based on classical dynamics although the parameters are fitted to non-classical DFT, and there is no zero-point motion in classical dynamics.

The author raises an open question. Is it ever possible for a classical MD to give a non-zero ADP at $T \rightarrow 0$ in systems where atoms are in a harmonic potential, such as in MgO?

## 5. Summary

The anisotropic ADPs by atom were directly derived from MD simulations taking the (co)variances over atom positions at different time steps. A universal machine-learned NNP was used to accelerate the calculations. The proposed method is applicable to systems with substitutional disorder and split sites, unlike with conventional methods using the dynamical matrix from phonon modes calculated at 0 K. The ADPs can be obtained by atom, where the (co)variance of each atom is obtained and then averaged for atoms in an atom site, or by site, where the (co)variance is calculated over all atoms at the atom site. The zero-point motion is not reflected in an ADP from classical MD simulations, but it can be estimated by extrapolation assuming an Einstein model using the proportionality factor between the ADP and temperature, if proportionality holds. The proposed method gives ADPs of MgO that are consistent with experimental values. The experimentally obtained shapes of anisotropic displacement ellipsoids of rattling atoms in thermoelectric materials $Ag_8SnSe_6$ and $Na_2In_2Sn_4$ are consistent with calculated results. The possibility of splitting of the Ag3 site of $Ag_8SnSe_6$ can be detected through calculation of the ADP over different temperatures. Migration of atoms was found in $Na_2In_2Sn_4$ at $T \geq 200$ K, resulting in a very large ADP by atom that disagrees with the experimentally obtained, much smaller ADPs by atom site. The ADPs by atom and by site is clearly different in the thermoelectric material $BaCu_{1.14}In_{0.86}P_2$, which is caused by disorder in the shared Cu/In site. The non-zero ADPs by site at $T \to 0$ in $BaCu_{1.14}In_{0.86}P_2$ is a good reason why ADPs should not be referred to as “temperature factors” or “thermal ellipsoids” because there are contributions to the 0 K ADPs from disorder of Cu/In atoms other than the zero-point motion. The investigations in this study suggest the effectiveness and limitations of direct ADP derivation from MD simulations and the use of calculated ADPs as a tool complementing experimental efforts to determine the crystal structure including atom displacement around atom sites.

**Acknowledgments**

The author acknowledges Dr. Naoya Ishida (AIST) for re-refinement of the published single crystal diffraction data of $BaCu_{1.14}In_{0.86}P_2$ (CCDC Number: 2285845)(Sarkar *et al.*, 2024a) based on the computationally suggested model. The author also thanks Dr. Eiji Nishibori for sharing experimental ADPs for $Ag_8SnSe_6$ used in Ref (Yamada *et al.*, 2023), which was used to draw Figs. 1(a) and 5(c) in this paper. This study was funded by a Kakenhi Grant-in-Aid (No. 24H00395) from the Japan Society for the Promotion of Science (JSPS). The VESTA code (Momma & Izumi, 2011) was used to draw Figure 1.

## Appendix. Mathematical details.

### A-1. Relations between different ADP definitions.

The **U**, **β**, and **B** are related by (equation 38 in ref. (Trueblood *et al.*, 1996))

$$U^{jl} = \beta^{jl}/(2\pi^2 a^j a^l) = B^{jl}/8\pi^2 \quad (5)$$

Using a scaling matrix

$$\mathbf{S} = \begin{pmatrix} a^* & 0 & 0 \\ 0 & b^* & 0 \\ 0 & 0 & c^* \end{pmatrix}, \tag{6}$$

$$\boldsymbol{\beta} = 2\pi^2 \mathbf{SUS} \,. \tag{7}$$

The Debye-Waller factor for a diffraction vector

$$\mathbf{h} = \sum_{i=1}^{3} h_i \mathbf{a}^i = \sum_{i=1}^{3} h_i^C \mathbf{e}_i \tag{8}$$

a row vector in reciprocal space, is

$$\begin{aligned} T(\mathbf{h}) &= \exp\left( -\sum_{j=1}^{3}\sum_{l=1}^{3} h_j \beta^{jl} h_l \right) \\ &= \exp\left( -2\pi^2 \sum_{j=1}^{3}\sum_{l=1}^{3} h_j a^j U^{jl} a^l h_l \right). \\ &= \exp\left( -2\pi^2 \sum_{j=1}^{3}\sum_{l=1}^{3} h_j^C U_{jl}^C h_l^C \right) \end{aligned} \tag{9}$$

(from equations 21, 24, 34, and 36 in ref. (Trueblood *et al.*, 1996)). Here, ($\mathbf{e}_1$, $\mathbf{e}_2$, $\mathbf{e}_3$) is an orthonormal basis.

One transformation matrix between the bases (**a**, **b**, **c)** and ($\mathbf{e}_1$, $\mathbf{e}_2$, $\mathbf{e}_3$) is (equation 50 in ref. (Trueblood *et al.*, 1996))

$$\mathbf{A} = \begin{pmatrix} a & b\cos\gamma & c\cos\beta \\ 0 & b\sin\beta & -c\sin\beta\cos\alpha^* \\ 0 & 0 & 1/c^* \end{pmatrix} \tag{10}$$

Coordinates of vector **r** in bases (**a**, **b**, **c**), ($a^*\mathbf{a}$, $b^*\mathbf{b}$, $c^*\mathbf{c}$), and ($\mathbf{e}_1$, $\mathbf{e}_2$, $\mathbf{e}_3$), which are denoted as ($x$, $y$, $z$), ($\xi$, $\eta$, $\zeta$), and ($\xi^C$, $\eta^C$, $\zeta^C$), respectively, transform as

$$(\xi^C, \eta^C, \zeta^C)^T = \mathbf{A}\,(x,y,z)^T \tag{11}$$

(based on equation 41 in ref. (Trueblood *et al.*, 1996)) and

$$((\xi^C, \eta^C, \zeta^C)^T = \mathbf{AS}(\xi, \eta, \zeta)^T \quad (12)$$

(based on equations 46-48 in ref. (Trueblood *et al.*, 1996)). Matrix

$$\mathbf{D} = \mathbf{AS} \quad (13)$$

was used in ref. (Trueblood *et al.*, 1996).

The transformation between **U** with the basis ($a^*\mathbf{a}$, $b^*\mathbf{b}$, $c^*\mathbf{c}$) and $\mathbf{U}^C$ with the orthonormal basis ($\mathbf{e}_1$, $\mathbf{e}_2$, $\mathbf{e}_3$) is

$$\mathbf{U}^C = \mathbf{ASUSA}^T = \mathbf{DUD}^T \quad (14)$$

(based on equation 49 in ref. (Trueblood *et al.*, 1996)).

The ADPs of an atom are often visualized as an ellipsoid. The principal semi-axes of the ellipsoid are simply the three eigenvalues of $\mathbf{U}^C$, and the directions of the semi-axes are those of the eigenvectors. The matrix of three eigenvectors **P** and the diagonal matrix with three corresponding eigenvalues **Q** are obtained as

$$\mathbf{P}^{-1}\mathbf{U}^C\mathbf{P} = \mathbf{Q} \quad (15)$$

The eigenvectors in the bases (**a**, **b**, **c**) and ($a^*\mathbf{a}$, $b^*\mathbf{b}$, $c^*\mathbf{c}$) are derived with equations 10 and 11, respectively. Alternatively, principal component analysis may be used to obtain the eigenvalues and eigenvectors of $\mathbf{U}^C$.

The conversion of **U** between two atoms with the same Wyckoff position, *p* and *q*, are given below. Let the mean positions of the two atoms *p* and *q*, $\mathbf{r} = (x, y, z)^T$ and $\mathbf{r}' = (x', y', z')^T$, be related as

$$\mathbf{r}' = \mathbf{Rr} + \mathbf{t} \quad (16)$$

where **R** is a rotation matrix and **t** is a translation vector. The relation between displacement vectors $\mathbf{u} = (\Delta x, \Delta y, \Delta z)^T$ and $\mathbf{u}' = (\Delta x', \Delta y', \Delta z')^T$ is

$$\mathbf{u}' = \mathbf{Ru}. \quad (17)$$

The goal here is to change the basis vectors defining $\mathbf{U}'$ such that the transformed quantity $\mathbf{U}''$ is directly comparable to **U**, which is accomplished by

$$\mathbf{u}' = \mathbf{R}\mathbf{u}\,. \tag{18}$$

This transformation is useful to increase the number of atoms with **U** values that can be easily averaged, thereby improving the reliability of results. The average **U** and $\mathbf{U}^C$ of atoms that are directly comparable is simply the mean of **U** and $\mathbf{U}^C$ of the atoms, respectively.

**A-2. Temperature dependence of the ADP in Einstein and Debye models**

The ADP in the Einstein model is discussed first. A 1D harmonic potential is considered with the standard Hamiltonian

$$\hat{H} = \left( -\frac{\hbar^2}{2m}\frac{d^2}{dx^2} + \frac{m\omega^2}{2}x^2 \right), \tag{19}$$

where $x$ is the position, $m$ is the particle mass, and ω is the characteristic angular frequency. An atom is placed in this potential. Although this potential does not have terms describing explicit interactions with other atoms, this potential reflects the interactions between atoms.

The eigenvalues of this potential are

$$E_n = \left(n + 1/2\right)\hbar\omega, \tag{20}$$

where $n$ are non-negative integers. The corresponding variance of the position, which is also the same as the ADP, $U$, is

$$U_n = \left\langle x^2 \right\rangle = \frac{E_n}{m\omega^2}\,. \tag{21}$$

In the classical limit of continuous $E_n$, and when the equipartition theorem holds,

$$U = \left\langle x^2 \right\rangle \propto E \propto T\,. \tag{22}$$

The partition function of the eigenvalues in equation 17 is

$$Z = \sum_{n=0}^{\infty} \exp\left( -\frac{E_n}{kT} \right) = \left\{ 2\sinh\left( \frac{\hbar\omega}{2kT} \right) \right\}^{-1}, \tag{23}$$

and $U$ for a certain $T$ is

$$U = Z^{-1}\sum_{n=0}^{\infty} U_n \exp\left(-\frac{E_n}{kT}\right) = \frac{\hbar}{2m\omega}\coth\left(\frac{\hbar\omega}{2kT}\right). \quad (24)$$

The limits of $U$ at $kT >> \hbar\omega$ and $kT << \hbar\omega$ are

$$U_{\mathrm{low}T} = \frac{\hbar}{2m\omega}. \quad (25)$$

and

$$U_{\mathrm{high}T} = \frac{kT}{m\omega^2} = U_{\mathrm{low}T}\frac{T}{T_{\mathrm{c}}}. \quad (26)$$

respectively, using a crossover temperature where $U_{\mathrm{low}T} = U_{\mathrm{high}T}$, namely

$$T_{\mathrm{c}} = \frac{\hbar\omega}{2k}. \quad (27)$$

When a set of $U$ and $T$ in the $U \propto T$ regime, $(U_1, T_1)$, is known,

$$\omega = \sqrt{\frac{kT_1}{mU_1}}. \quad (28)$$

resulting in

$$U_{\mathrm{low}T} = \frac{\hbar}{2}\sqrt{\frac{U_1}{mkT_1}} = 3.4824\sqrt{\frac{U_1}{mT_1}}. \quad (29)$$

$$T_{\mathrm{c}} = \frac{\hbar}{2}\sqrt{\frac{T_1}{mkU_1}} = 3.4824\sqrt{\frac{T_1}{mU_1}}. \quad (30)$$

and

$$U = U_{\mathrm{low}T}\coth\left(T_{\mathrm{c}}/T\right). \quad (31)$$

The crossover temperature tends to be higher in light atoms with small $U/T$ atoms. Atoms with smaller $U$ are more tightly bound to the equilibrium point.

The rightmost term in equations 29 and 30 is when units of daltons for mass, K for temperature, and Å$^2$ for $U$ are used. The proportionality factor between $U$ and $T$ may be used instead of $U_1/T_1$ if this can be obtained from fitting.

The correction factor to $U$ arising from the zero-point motion may be written as

$$\frac{U - U_{\mathrm{high}T}}{U_{\mathrm{high}T}} = \frac{T_{\mathrm{c}}}{T}\coth\left(\frac{T_{\mathrm{c}}}{T}\right) - 1. \quad (32)$$

The function $f\left(x\right)=\coth\left(1/x\right)/x-1$ converges relatively slowly to 0 when $x$ is increased. Its value when $x$ = 1, 2, 3, 4, 5, 6, 10 is 0.313, 0.082, 0.037, 0.021, 0.013, 0.019, and 0.003, respectively. In other words, the correction factor is 30%, 10%, 3%, and 1% when $T/T_c$ is 1.023, 1.81, 3.3, and 5.8, respectively.

The Debye model is considered next. The simplest model of a 3D crystal is considered with one atom per unit cell, degenerate acoustic branches, and polarization vectors forming an orthonormal tern is considered. (Grosso & Parravicini, 2014) The $U$ from the Debye model along a certain direction is, using a Debye temperature $T_D$,

$$U_{\mathrm{D}}=3\frac{\hbar^2T^2}{mkT_{\mathrm{D}}^3}\int_0^{T_{\mathrm{D}}/T}\left(\frac{1}{e^x-1}+\frac{1}{2}\right)xdx\,. \tag{33}$$

The low and high temperature approximations are

$$U_{\mathrm{D_low}T}=\frac{3}{4}\frac{\hbar^2}{mkT_{\mathrm{D}}} \tag{34}$$

and

$$U_{\mathrm{D_high}T}=3\frac{\hbar^2T}{mkT_{\mathrm{D}}^2}\,, \tag{35}$$

respectively. The crossover temperature where $U_{\mathrm{D_low}T}=U_{\mathrm{D_high}T}$ is

$$T_{\mathrm{D_c}}=T_{\mathrm{D}}/4\,. \tag{36}$$

The $U_{\mathrm{D_low}T}$ can be obtained from a set of $U$ and $T$ in the $U\propto T$ regime, ($U_1$, $T_1$), as

$$U_{\mathrm{D_low}T}=\frac{\hbar}{4}\sqrt{\frac{3U_1}{mkT_1}}=3.0159\sqrt{\frac{U_1}{mT_1}}\,. \tag{37}$$

The rightmost term is when units of daltons for mass, K for temperature, and Å$^2$ for $U$ are used. According to equations 30 and 37,

$$U_{\mathrm{D_low}T}=\left(\sqrt{3}/2\right)U_{\mathrm{low}T} \tag{38}$$

and

$$T_{\mathrm{D_c}}=\left(\sqrt{3}/2\right)T_{\mathrm{c}}\,. \tag{35}$$

Note that the $U_{\mathrm{D_low}T}$ must be the same for all atoms and all directions in this simplest Debye model.

**References**

Beg, M. (1976). *Acta Crystallographica Section A* **32**, 154-156.

Erba, A., Ferrabone, M., Orlando, R. & Dovesi, R. (2013). *J. Comput. Chem.* **34**, 346-354.

Grosse-Kunstleve, R. W. & Adams, P. D. (2002). *J. Appl. Crystallogr.* **35**, 477-480.

Grosso, G. & Parravicini, G. P. (2014). *Solid State Physics (Second Edition)*, edited by G. Grosso & G. P. Parravicini, pp. 391-436. Amsterdam: Academic Press.

Hinuma, Y. (2025). *Comp. Mater. Sci.* **246**, 113368.

Hoover, W. G. (1985). *Phys. Rev. A* **31**, 1695-1697.

Iversen, B. B., Larsen, F. K., Figgis, B. N., Reynolds, P. A. & Schultz, A. J. (1996). *Acta Cryst. B* **52**, 923-931.

Jacobs, R., Morgan, D., Attarian, S., Meng, J., Shen, C., Wu, Z., Xie, C. Y., Yang, J. H., Artrith, N., Blaiszik, B., Ceder, G., Choudhary, K., Csanyi, G., Cubuk, E. D., Deng, B., Drautz, R., Fu, X., Godwin, J., Honavar, V., Isayev, O., Johansson, A., Kozinsky, B., Martiniani, S., Ong, S. P., Poltavsky, I., Schmidt, K. J., Takamoto, S., Thompson, A. P., Westermayr, J. & Wood, B. M. (2025). *Curr. Opin. Solid State Mater. Sci.* **35**, 101214.

Johnson, C. K. (1965). ORTEP: A FORTRAN Thermal-Ellipsoid Plot Program for Crystal Structure Illustrations, ORNL-3794 (Rev.), Union Carbide Corp., Oak Ridge Natl. Lab, June 1965.

Kresse, G. & Furthmüller, J. (1996). *Phys. Rev. B* **54**, 11169-11186.

Kresse, G. & Joubert, D. (1999). *Phys. Rev. B* **59**, 1758-1775.

Lane, N. J., Vogel, S. C., Hug, G., Togo, A., Chaput, L., Hultman, L. & Barsoum, M. W. (2012). *Phys. Rev. B* **86**, 214301.

Lawrence, J. L. (1973). *Acta Crystallographica Section A* **29**, 94-95.

Madsen, A. O., Civalleri, B., Ferrabone, M., Pascale, F. & Erba, A. (2013). *Acta Crystallographica Section A* **69**, 309-321.

Momma, K. & Izumi, F. (2011). *J. Appl. Crystallogr.* **44**, 1272-1276.

Nosé, S. (1984). *J. Chem. Phys.* **81**, 511-519.

Perdew, J. P., Burke, K. & Ernzerhof, M. (1996). *Phys. Rev. Lett.* **77**, 3865-3868.

Reilly, A. M., Wann, D. A., Gutmann, M. J., Jura, M., Morrison, C. A. & Rankin, D. W. H. (2013). *J. Appl. Crystallogr.* **46**, 656-662.

Reilly, A. M., Wann, D. A., Morrison, C. A. & Rankin, D. W. H. (2007). *Chem. Phys. Lett.* **448**, 61-64.

Sarkar, A., Porter, A. P., Viswanathan, G., Yox, P., Earnest, R. A., Wang, J., Rossini, A. J.

& Kovnir, K. (2024a) CSD 2285845: Experimental Crystal Structure Determination https://doi.org/10.1039/d4ta01063a.

Sarkar, A., Porter, A. P., Viswanathan, G., Yox, P., Earnest, R. A., Wang, J., Rossini, A. J. & Kovnir, K. (2024b). *Journal of Materials Chemistry A* **12**, 10481-10493.

Sasaki, S., Fujino, K., Tak, Eacute & Uchi, Y. (1979). *Proceedings of the Japan Academy, Series B* **55**, 43-48.

Suekuni, K., Lee, C. H., Tanaka, H. I., Nishibori, E., Nakamura, A., Kasai, H., Mori, H., Usui, H., Ochi, M., Hasegawa, T., Nakamura, M., Ohira-Kawamura, S., Kikuchi, T., Kaneko, K., Nishiate, H., Hashikuni, K., Kosaka, Y., Kuroki, K. & Takabatake, T. (2018). *Adv. Mater.* **30**, 1706230.

Takahashi, S., Kasai, H., Liu, C., Miao, L. & Nishibori, E. (2024). *Crystal Growth & Design* **24**, 6267-6274.

Takamoto, S., Shinagawa, C., Motoki, D., Nakago, K., Li, W., Kurata, I., Watanabe, T., Yayama, Y., Iriguchi, H., Asano, Y., Onodera, T., Ishii, T., Kudo, T., Ono, H., Sawada, R., Ishitani, R., Ong, M., Yamaguchi, T., Kataoka, T., Hayashi, A., Charoenphakdee, N. & Ibuka, T. (2022). *Nat. Commun.* **13**, 2991.

Trueblood, K. N., Burgi, H.-B., Burzlaff, H., Dunitz, J. D., Gramaccioli, C. M., Schulz, H. H., Shmueli, U. & Abrahams, S. C. (1996). *Acta Crystallographica Section A* **52**, 770-781.

Tsirelson, V. G., Avilov, A. S., Abramov, Y. A., Belokoneva, E. L., Kitaneh, R. & Feil, D. (1998). *Acta Cryst. B* **54**, 8-17.

White, G. K. & Anderson, O. L. (1966). *J. Appl. Phys.* **37**, 430-432.

Whitten, A. E. & Spackman, M. A. (2006). *Acta Cryst. B* **62**, 875-888.

Willis, B. T. M. & Pryor, A. W. (1975). *Thermal Vibrations in Crystallography*. London, UK: Cambridge University Press.

Yamada, T., Yoshiya, M., Kanno, M., Takatsu, H., Ikeda, T., Nagai, H., Yamane, H. & Kageyama, H. (2023). *Adv. Mater.* **35**, 2207646.

**Table 1.** Notations of position and displacement vectors, bases, and ADPs.

| Basis | Direct lattice (**a**, **b**, **c**) or ($\mathbf{a}_1$, $\mathbf{a}_2$, $\mathbf{a}_3$) | Direct lattice ($a^*\mathbf{a}$, $b^*\mathbf{b}$, $c^*\mathbf{c}$) or ($a^1\mathbf{a}_1$, $a^2\mathbf{a}_2$, $a^3\mathbf{a}_3$) | Cartesian basis ($\mathbf{e}_1$, $\mathbf{e}_2$, $\mathbf{e}_3$) |
|---|---|---|---|
| Components of **r** | $x, y, z$ or $x^1, x^2, x^3$ | $\xi, \eta, \zeta$ or $\xi^1, \xi^2, \xi^3$ | $\xi^C, \eta^C, \zeta^C$ or $\xi^C{}_1, \xi^C{}_2, \xi^C{}_3$ |
| Components of **u** | $\Delta x, \Delta y, \Delta z$ or $\Delta x^1, \Delta x^2, \Delta x^3$ | $\Delta\xi, \Delta\eta, \Delta\zeta$ or $\Delta\xi^1, \Delta\xi^2, \Delta\xi^3$ | $\Delta\xi^C, \Delta\eta^C, \Delta\zeta^C$ or $\Delta\xi^C{}_1, \Delta\xi^C{}_2, \Delta\xi^C{}_3$ |
| Related ADP | $\beta^{ij}=2\pi^2<\Delta x^i \Delta x^j>$ | $U^{ij}=<\Delta\xi^i \Delta\xi^j>$ | $U^C{}_{ij}=<\Delta\xi^C{}_i \Delta\xi^C{}_j>$ |
| Unit of related ADP | Dimensionless | Length$^2$ | Length$^2$ |

**Table 2.** **U** of Ag in $Ag_8SnSe_6$ at 200 K from 3×3×3 supercell MD simulations. The unit is Å$^2$.

| | $U^{11}$ | $U^{22}$ | $U^{33}$ | $U^{23}$ | $U^{13}$ | $U^{12}$ | $U_3$ | $U_2$ | $U_1$ | $U_3/U_1$ | $U_{iso}$ |
|---|---|---|---|---|---|---|---|---|---|---|---|
| Ag1 | 0.024 | 0.023 | 0.029 | 0.008 | 0.006 | 0.010 | 0.041 | 0.022 | 0.013 | 3.2 | 0.025 |
| Ag2 | 0.025 | 0.028 | 0.038 | -0.004 | 0.006 | -0.001 | 0.042 | 0.026 | 0.022 | 1.9 | 0.030 |
| Ag3 | 0.041 | 0.065 | 0.047 | 0.004 | -0.025 | 0.009 | 0.072 | 0.065 | 0.017 | 4.3 | 0.051 |
| Ag4 | 0.031 | 0.028 | 0.019 | -0.004 | 0.000 | 0.000 | 0.031 | 0.029 | 0.018 | 1.7 | 0.026 |
| Ag5 | 0.037 | 0.023 | 0.013 | 0.005 | 0.000 | 0.000 | 0.037 | 0.025 | 0.012 | 3.2 | 0.025 |

**Table 3.** **U** values of Ag in $Ag_8SnSe_6$ at 300 K from 3×3×3 supercell MD simulations. The unit is Å$^2$.

| | $U^{11}$ | $U^{22}$ | $U^{33}$ | $U^{23}$ | $U^{13}$ | $U^{12}$ | $U_3$ | $U_2$ | $U_1$ | $U_3/U_1$ | $U_{iso}$ |
|---|---|---|---|---|---|---|---|---|---|---|---|
| Ag1 | 0.042 | 0.040 | 0.053 | 0.015 | 0.013 | 0.019 | 0.076 | 0.037 | 0.022 | 3.5 | 0.045 |
| Ag2 | 0.043 | 0.051 | 0.066 | -0.014 | 0.003 | 0.002 | 0.075 | 0.046 | 0.040 | 1.9 | 0.053 |
| Ag3 | 0.060 | 0.110 | 0.068 | 0.010 | -0.034 | 0.014 | 0.114 | 0.098 | 0.026 | 4.3 | 0.080 |
| Ag4 | 0.052 | 0.048 | 0.034 | -0.008 | 0.000 | -0.001 | 0.053 | 0.051 | 0.031 | 1.7 | 0.045 |
| Ag5 | 0.065 | 0.038 | 0.022 | 0.007 | 0.000 | -0.001 | 0.065 | 0.040 | 0.019 | 3.3 | 0.041 |

**Table 4.** **U** of Na in $Na_2In_2Sn_4$ derived from MD simulations at various temperatures *T*. Experimental values from Yamada et al. (Yamada *et al.*, 2023) are shown in brackets. Units of *T* and **U** are K and Å$^2$, respectively.

| *T* | $U^{11}$ | $U^{22}$ | $U^{33}$ | $U^{23}$ | $U^{13}$ | $U^{12}$ |
|---|---|---|---|---|---|---|
| 50 | 0.009 | 0.009 | 0.009 | -0.004 | -0.005 | 0.005 |
| 100 | 0.020 | 0.021 | 0.018 | -0.009 | -0.011 | 0.011 |
| 150 | 0.033 | 0.033 | 0.028 | -0.015 | -0.017 | 0.019 |
| 200 | 0.046 | 0.047 | 0.039 | -0.021 | -0.024 | 0.030 |
| | (0.055) | (0.051) | (0.048) | (-0.024) | (-0.033) | (0.036) |
| 250 | 0.072 | 0.070 | 0.065 | -0.029 | -0.036 | 0.037 |
| | (0.072) | (0.065) | (0.061) | (0.049) | (-0.043) | (-0.032) |
| 300 | 0.092 | 0.088 | 0.094 | -0.029 | -0.051 | 0.034 |
| | (0.087) | (0.078) | (0.073) | (-0.040) | (0.055) | (0.060) |

**Table 5.** **U** of $BaCu_{1.14}In_{0.86}P_2$ at 175 K (Sarkar *et al.*, 2024b). Values from MD simulations are shown together with experimental values from Sarkar et al. (Sarkar *et al.*, 2024b) **U** values corrected for the zero-point motion based on the Einstein model are also shown. In Sarkar et al. (Sarkar *et al.*, 2024b), the $U_{iso}$ of Ba is the same for Ba1 and Ba11 sites and the ADP of Ba is not given, and only the combined ADP is provided for Cu/In sites. The Cu/In values from calculations, shown in brackets, are the 164:124 weighted averages based on the Cu/In atom ratio. The unit of **U** is $10^{-4}$Å$^2$.

| | Calculated | | | | Corrected | | | | Experiment | | | |
|---|---|---|---|---|---|---|---|---|---|---|---|---|
| | $U^{11}$ | $U^{33}$ | $U^{33}/U^{11}$ | $U_{iso}$ | $U^{11}$ | $U^{33}$ | $U^{33}/U^{11}$ | $U_{iso}$ | $U^{11}$ | $U^{33}$ | $U^{33}/U^{11}$ | $U_{iso}$ |
| Ba | 77 | 101 | 1.31 | 85 | 80 | 104 | 1.31 | 88 | | | | 103 |
| Cu | 116 | 193 | 1.66 | 142 | 121 | 197 | 1.63 | 147 | | | | |
| In | 65 | 71 | 1.09 | 67 | 68 | 74 | 1.08 | 70 | | | | |
| Cu/In | (94) | (140) | (1.49) | (110) | (99) | (144) | (1.46) | (114) | 165 | 222 | 1.35 | 184 |
| P | 70 | 71 | 1.01 | 70 | 77 | 78 | 1.01 | 78 | 93 | 208 | 2.24 | 131 |

**Table 6.** Mean average and standard deviation (std. dev.) over 12 temperatures of **U**/$T$, by atom, of $BaCu_{1.14}In_{0.86}P_2$ as well as $U_{lowT}$ and $T_c$ based on the Einstein model. The (standard deviation) / (mean average) over all considered temperatures was 4.6% or less for all of $U^{11}$, $U^{33}$, and $U_{iso}$ of all elements.

| | Mean ($10^{-5}$Å$^2$/K) | | | Std. dev. ($10^{-5}$Å$^2$/K) | | | $U_{lowT}$ ($10^{-4}$Å$^2$) | | | $T_c$ (K) | | |
|---|---|---|---|---|---|---|---|---|---|---|---|---|
| | $U^{11}$ | $U^{33}$ | $U_{iso}$ | $U^{11}$ | $U^{33}$ | $U_{iso}$ | $U^{11}$ | $U^{33}$ | $U_{iso}$ | $U^{11}$ | $U^{33}$ | $U_{iso}$ |
| Ba | 4.45 | 5.86 | 4.92 | 0.15 | 0.19 | 0.17 | 20 | 23 | 21 | 45 | 39 | 42 |
| Cu | 6.73 | 11.07 | 8.18 | 0.12 | 0.20 | 0.12 | 36 | 46 | 40 | 53 | 42 | 48 |
| In | 3.78 | 4.10 | 3.89 | 0.07 | 0.07 | 0.06 | 20 | 21 | 20 | 53 | 51 | 52 |
| Ba | 4.00 | 4.06 | 4.02 | 0.03 | 0.04 | 0.02 | 40 | 40 | 40 | 99 | 98 | 99 |

**Table 7.** Parameters used to fit curves in Fig. 9(e,f) using equation 4.

| Elements | $T$ (K) | $A$ ($10^3$) | $p$ | σ | Δ$z$ |
|---|---|---|---|---|---|
| Cu | 175 | 28.4 | 0.66 | 0.0115 | 0.0257 |
| Cu | 50 | 35.1 | 0.63 | 0.0092 | 0.0257 |
| Cu/In | 175 | 56.3 | 0.81 | 0.0101 | 0.0264 |
| Cu/In | 50 | 67.4 | 0.79 | 0.0084 | 0.0257 |

**Table 8.** Re-refinement of single crystal XRD data of $BaCu_{1.14}In_{0.86}P_2$ (Sarkar *et al.*, 2024b) based on a model suggested from MD simulations. Lattice parameters are $a$=$b$=4.0773(4) Å and $c$= 13.451(2) Å, "Occ." is the site occupancy factor.

| | $x$ | $y$ | $z$ | Occ. | $U^{11}$ (Å$^2$) | $U^{33}$ (Å$^2$) |
|---|---|---|---|---|---|---|
| Ba1 | 0 | 0 | 0 | 1 | 0.0106(3) | 0.0222(5) |
| Cu1 | 0.5 | 0 | 0.25 | 0.381(14) | 0.0163(4) | 0.0123(12) |
| Cu2 | 0.5 | 0 | 0.2720(17) | 0.096(7) | 0.0163(4) | 0.0123(12) |
| In1 | 0.5 | 0 | 0.25 | 0.427 | 0.0163(4) | 0.0123(12) |
| P1 | 0.5 | 0.5 | -0.1330(2) | 1 | 0.0093(7) | 0.0210(12) |

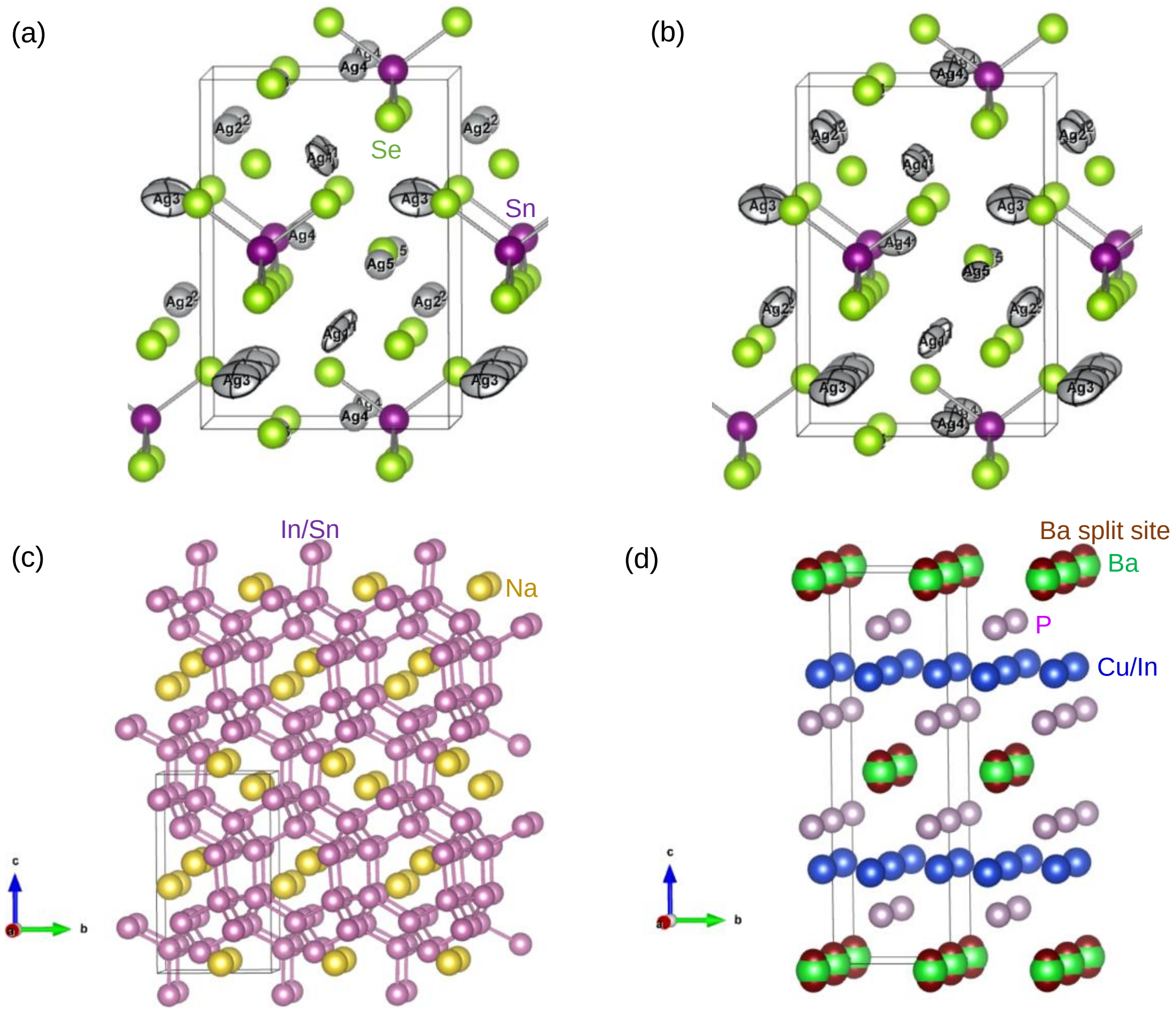

Fig. 1. Experimentally obtained crystal structures of (a,b) $Ag_8SnSe_6$ (Takahashi *et al.*, 2024), (c) $Na_2In_2Sn_4$ (Yamada et al., 2023), and (d) $BaCu_{1.14}Cu_{0.86}P_2$ (Sarkar *et al.*, 2024b). The 300 K experimentally obtained anisotropic ADPs for Ag1 and Ag3 are shown in (a), while the 300 K computationally derived ADPs for all Ag are given in (b). (a,b) Gray, dark purple, and green circles represent Ag, Sn, and Se sites respectively. (c) Yellow and purple circles represent Na and In/Sn sites, respectively. (d) Green, blue, and light purple circles represent the main Ba, Cu/In, and P sites, respectively. The small brown circles above and below large green circles are the Ba11 sites in ref. (Sarkar *et al.*, 2024b). This reference claims that the Ba site is triple-split where most of the Ba occupies the main Ba1 site (green circles) but about 18% enters Ba11 sites at 175 K.

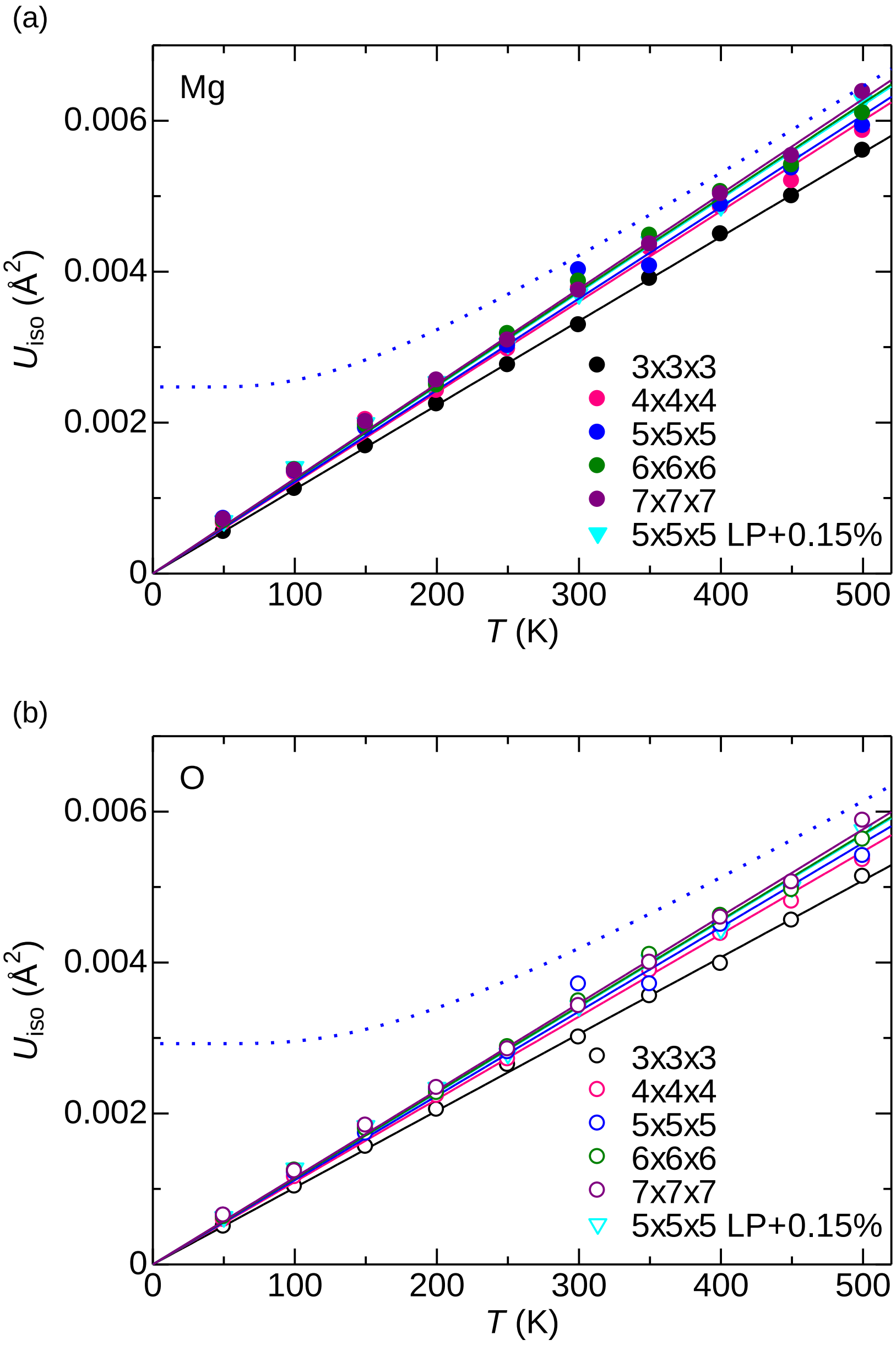

Fig. 2. ADP $U_{iso}$ versus temperature $T$ of (a) Mg and (b) O in MgO. The linear regressions pass through the origin. The dashed lines represent $U_{iso}$ of the 5×5×5 supercell corrected with the Einstein model. The LP+0.15% points are from the 5×5×5 supercell with the lattice parameter increased by 0.15% to account for the lattice expansion at 300 K.

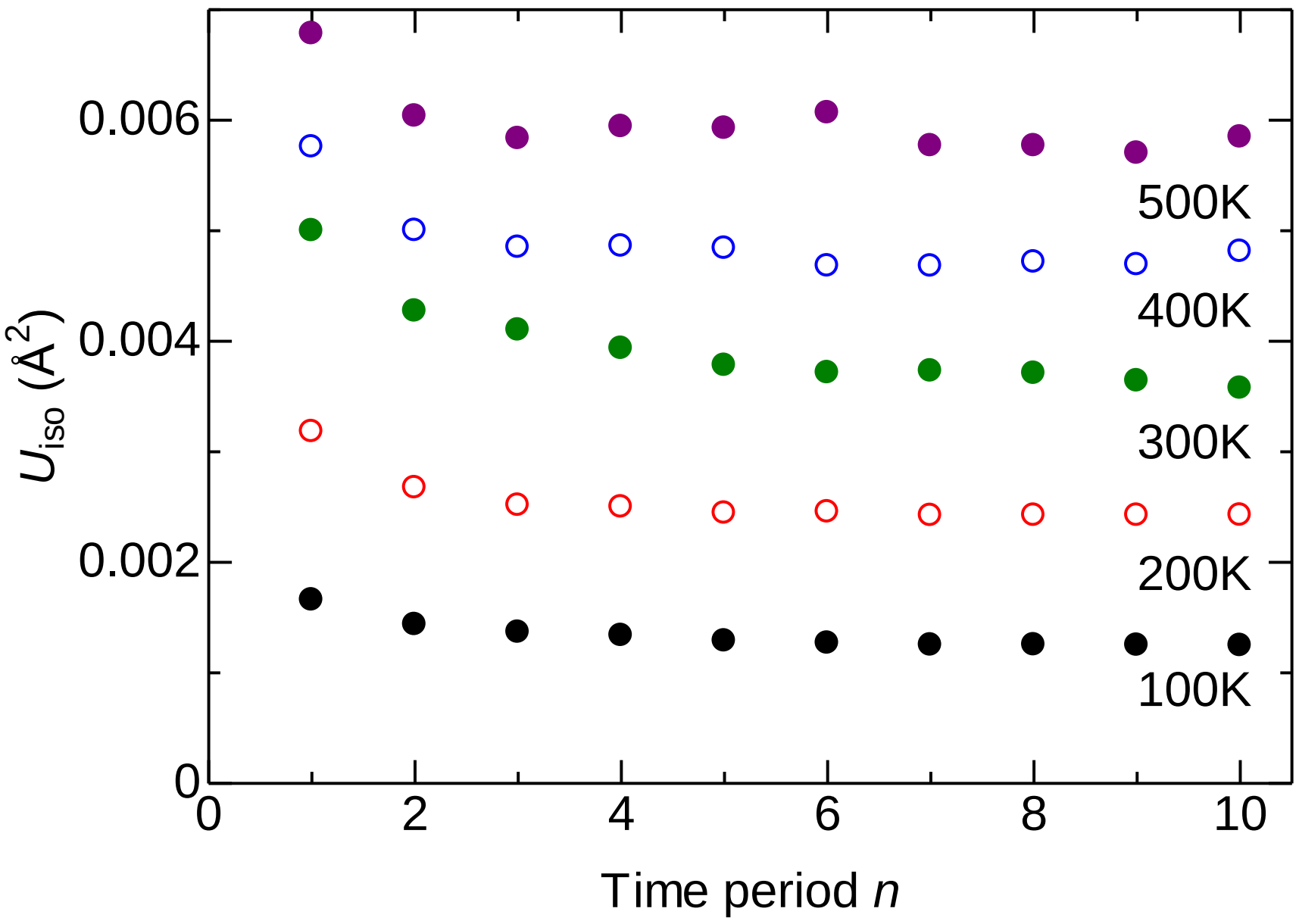

Fig. 3. $U_{iso}$ for MgO based on the $50(n\text{-}1)+1$ -th to $50n$ -th position recordings within the 500 position recordings from a MD simulation.

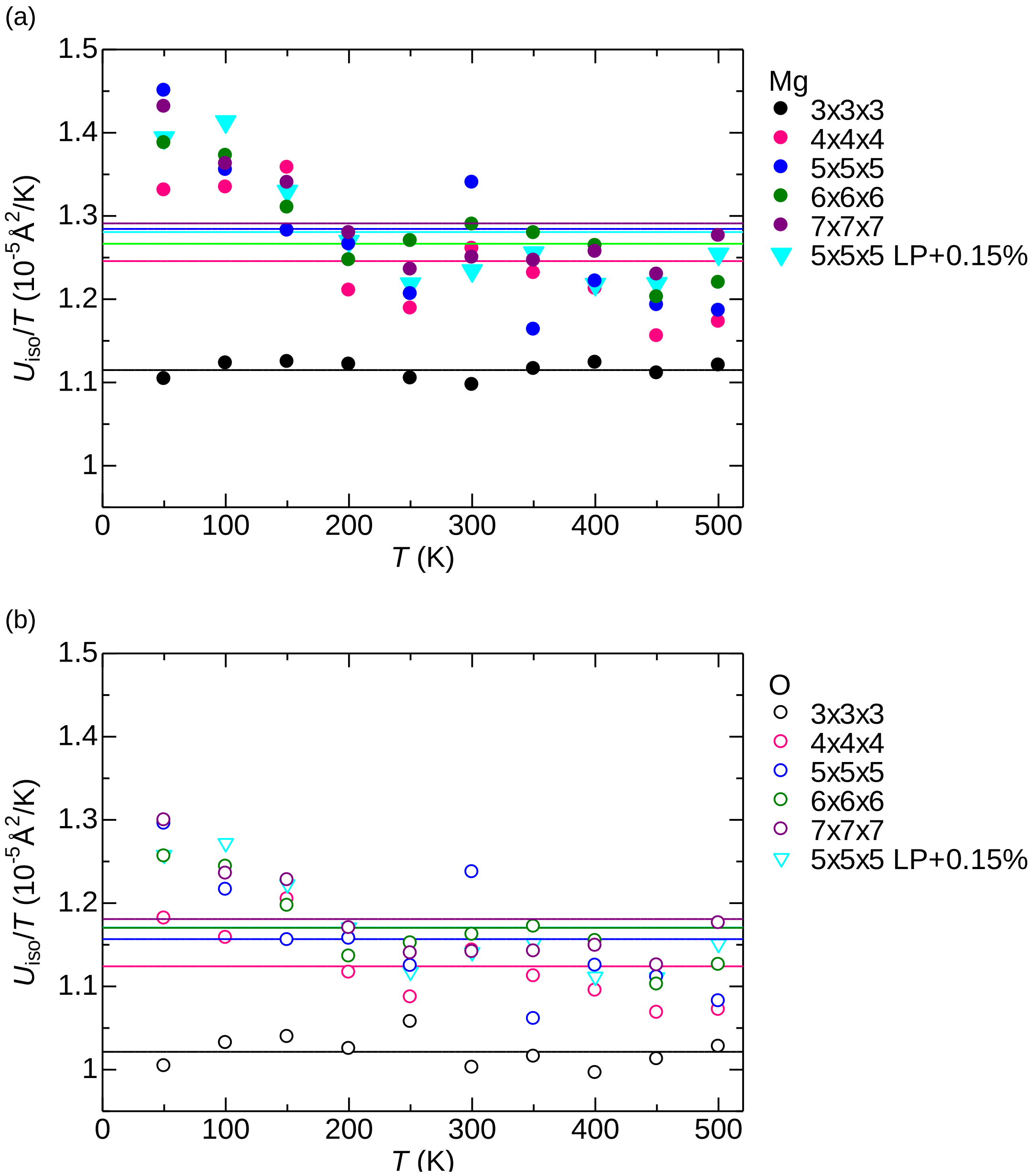

Fig. 4. ADP $U_{iso}$ divided by temperature $T$ versus $T$ for (a) Mg and (b) O in MgO. The horizontal lines are the average value. The LP+0.15% points are from the 5×5×5 supercell with the lattice parameter increased by 0.15% to account for the lattice expansion at 300 K. The horizontal lines for 6×6×6 and 5×5×5 LP+0.15% lines almost overlap.

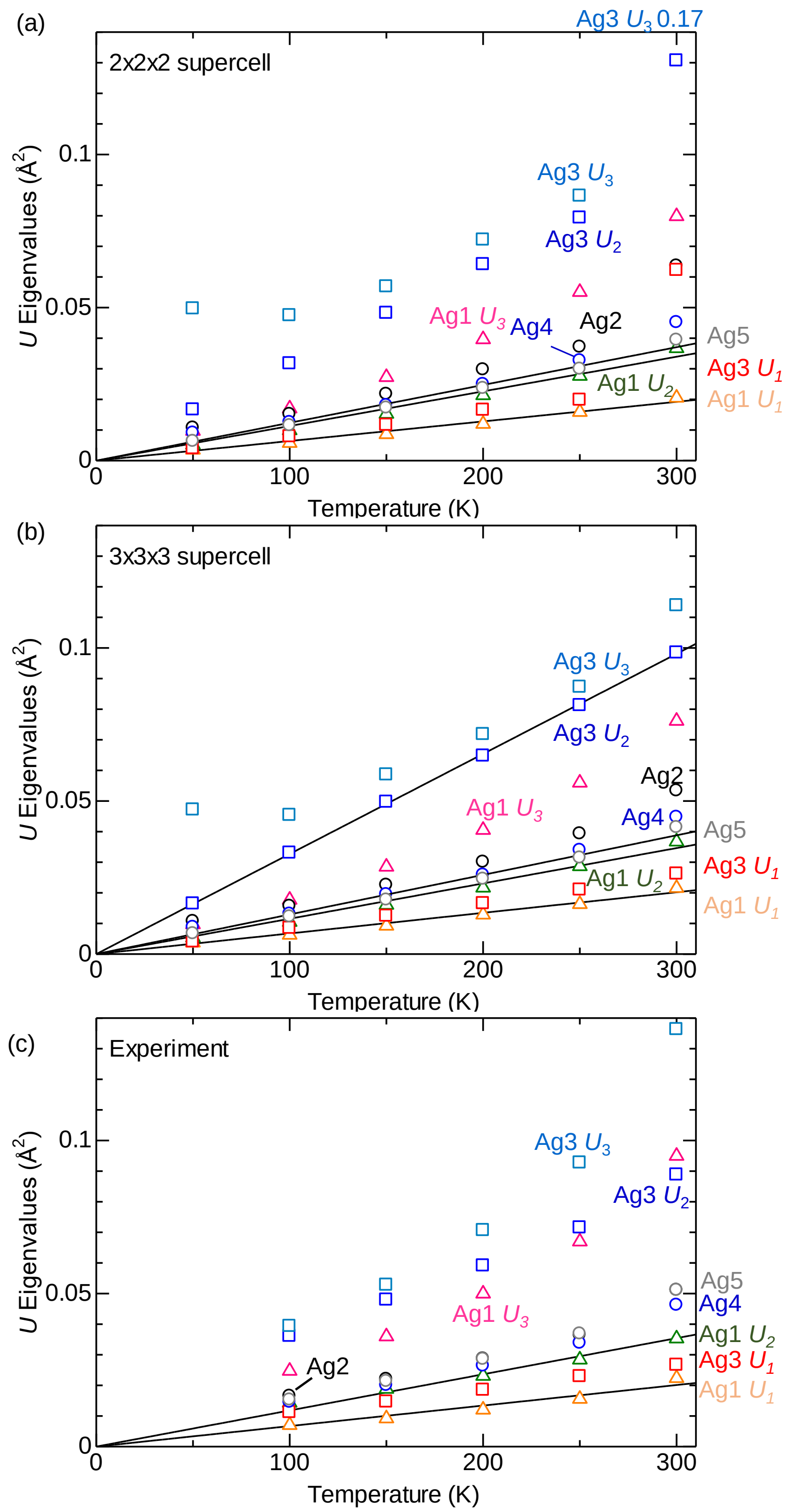

Fig. 5. (a,b) Computational and (c) experimental **U** values of $Ag_8InSe_6$. The displayed quantities and their symbols are the same as experimental data (shown in (c)) in Ref. (Yamada *et al.*, 2023). Linear regressions pass through the origin. A point in (a) of Ag3 $U_3$ at 0.17 is not shown in this vertical scale.

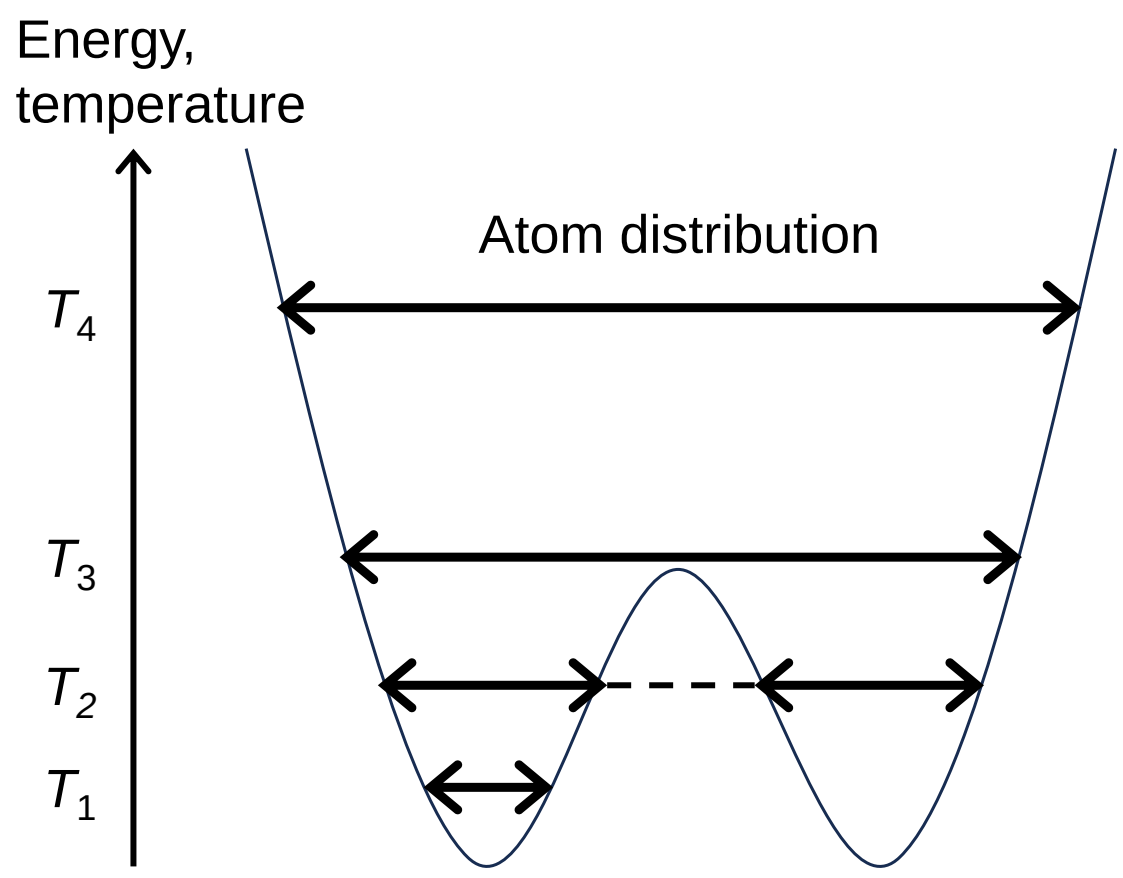

Fig. 6. Schematic of atom distribution in a double well potential.

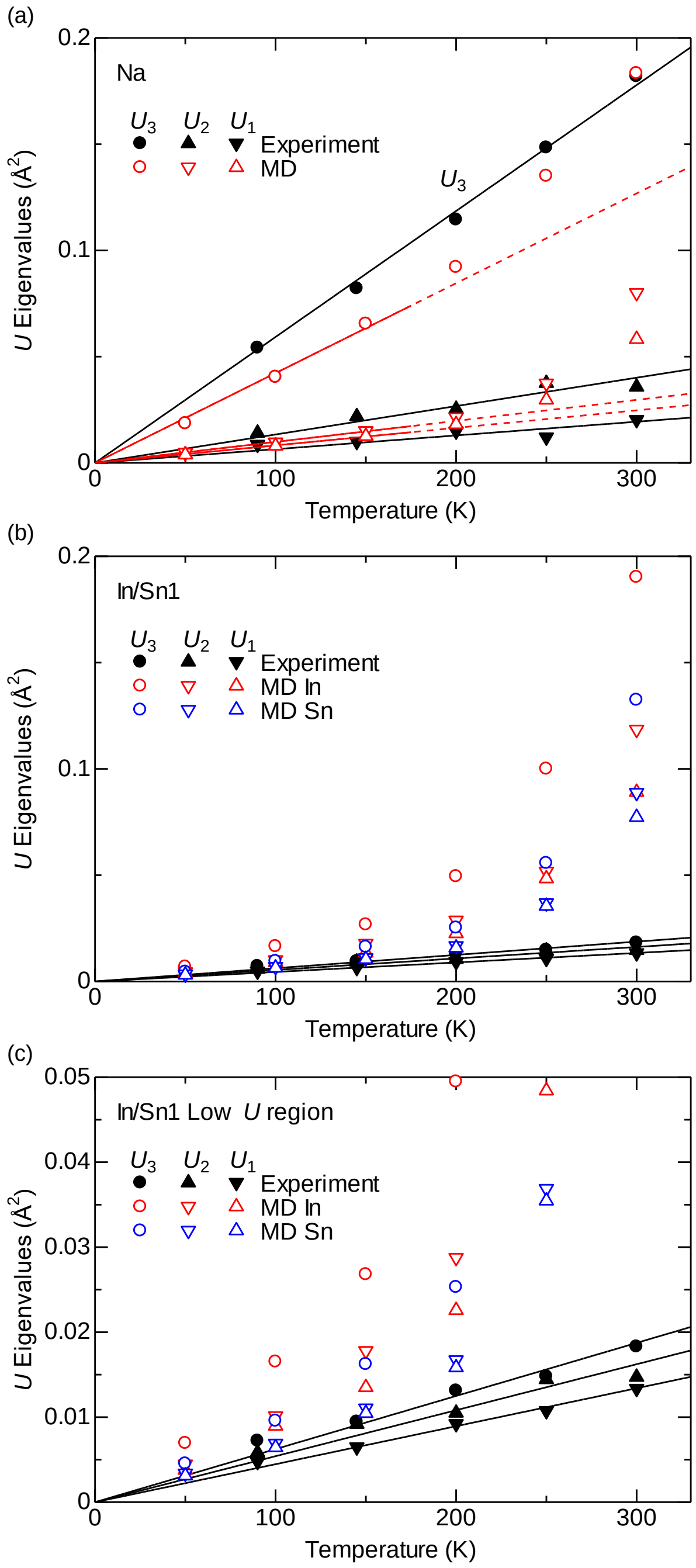

Fig. 7. The eigenvalues of experimental (Yamada *et al.*, 2023) and calculated **U** of $Na_2In_2Sn_4$, of (a) Na and (b.c) In/Sn sites, shown with different symbols, plotted against temperature. (c) is an enlargement of (b) for small **U**. Experimental values are shown with black filled symbols and the linear regressions passing through the origin are shown with solid lines. Computational values are shown with empty symbols and the linear regressions in (a) passing through the origin for 50, 100, and 150 K points are shown with solid lines at $T < 175$ K and are extrapolated using dashed lines at $T > 175$ K.

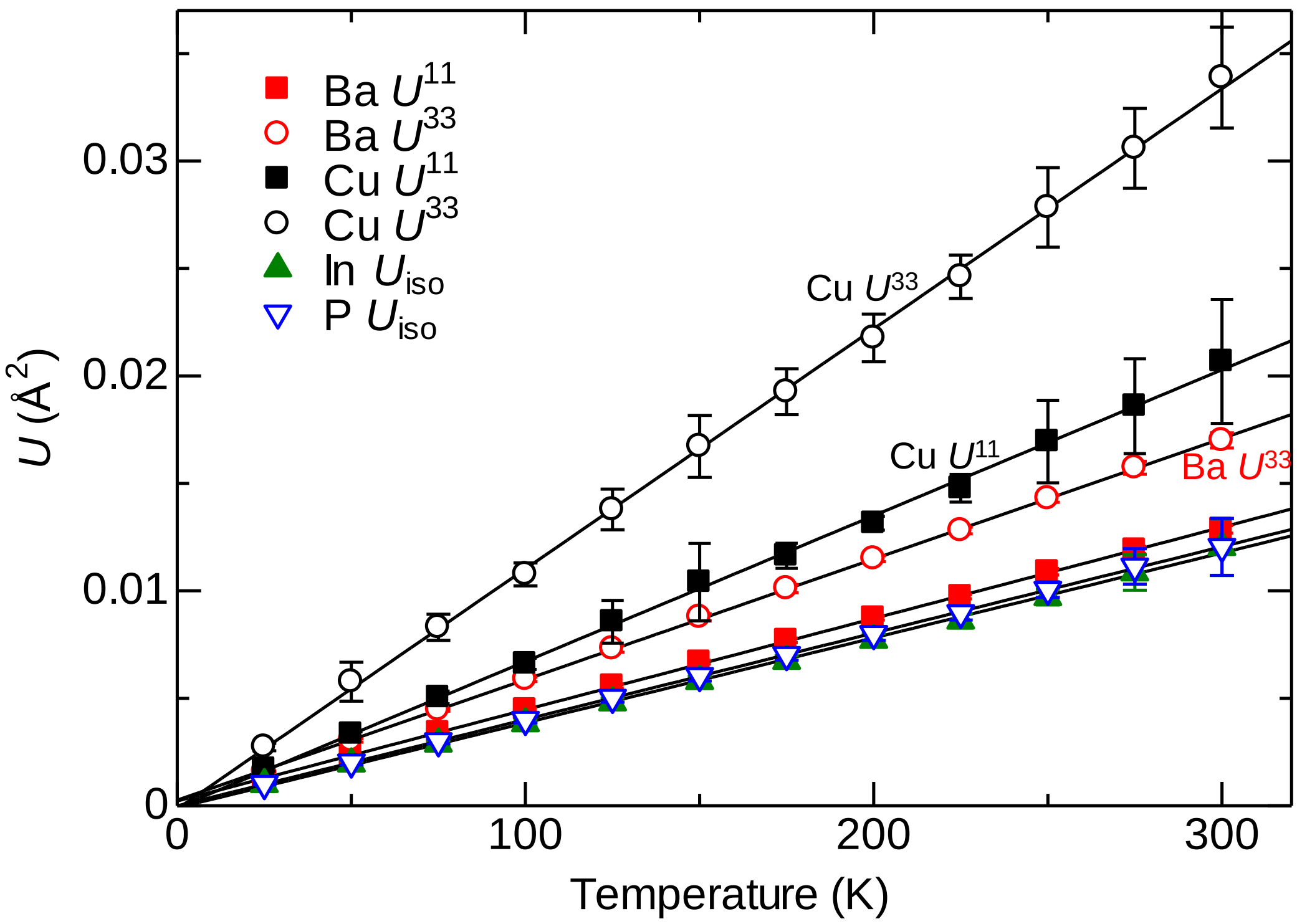

Fig. 8. Calculated ***U*** values by atom for $BaCu_{1.14}In_{0.86}P_2$. The anisotropic $U^{11}$ and $U^{33}$ are given for Ba and Cu because of their large anisotropy, and isotropic $U_{iso}$ is plotted for In and P with low anisotropy. The linear regressions in solid lines are not forced to pass through the origin. The range of ±1 standard deviation of **U** over 10 calculations are shown.

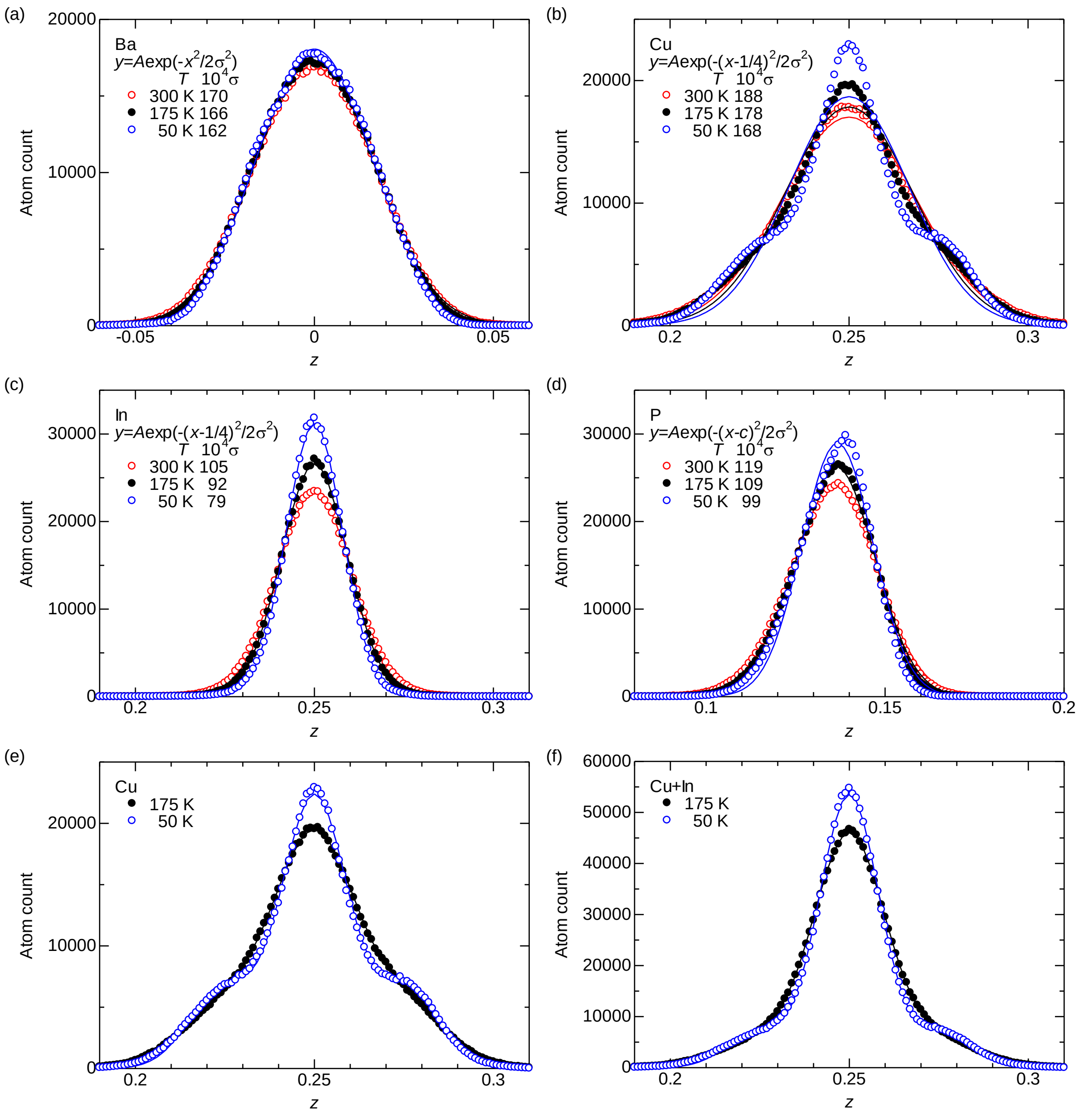

Fig. 9. Distribution of $z$ coordinates of (a) Ba, (b,e) Cu, (c) In, (d) P, and (f) Cu+In combined in $BaCu_{1.14}In_{0.86}P_2$ derived from the atom positions obtained from MD simulations. The atom positions were binned with 0.001 intervals of $z$. The points are fitted to (a-d) normal distributions (solid lines) and (e,f) superimposition of three normal distributions in equation 4. The standard deviation σ of the distributions are shown in (a-d).

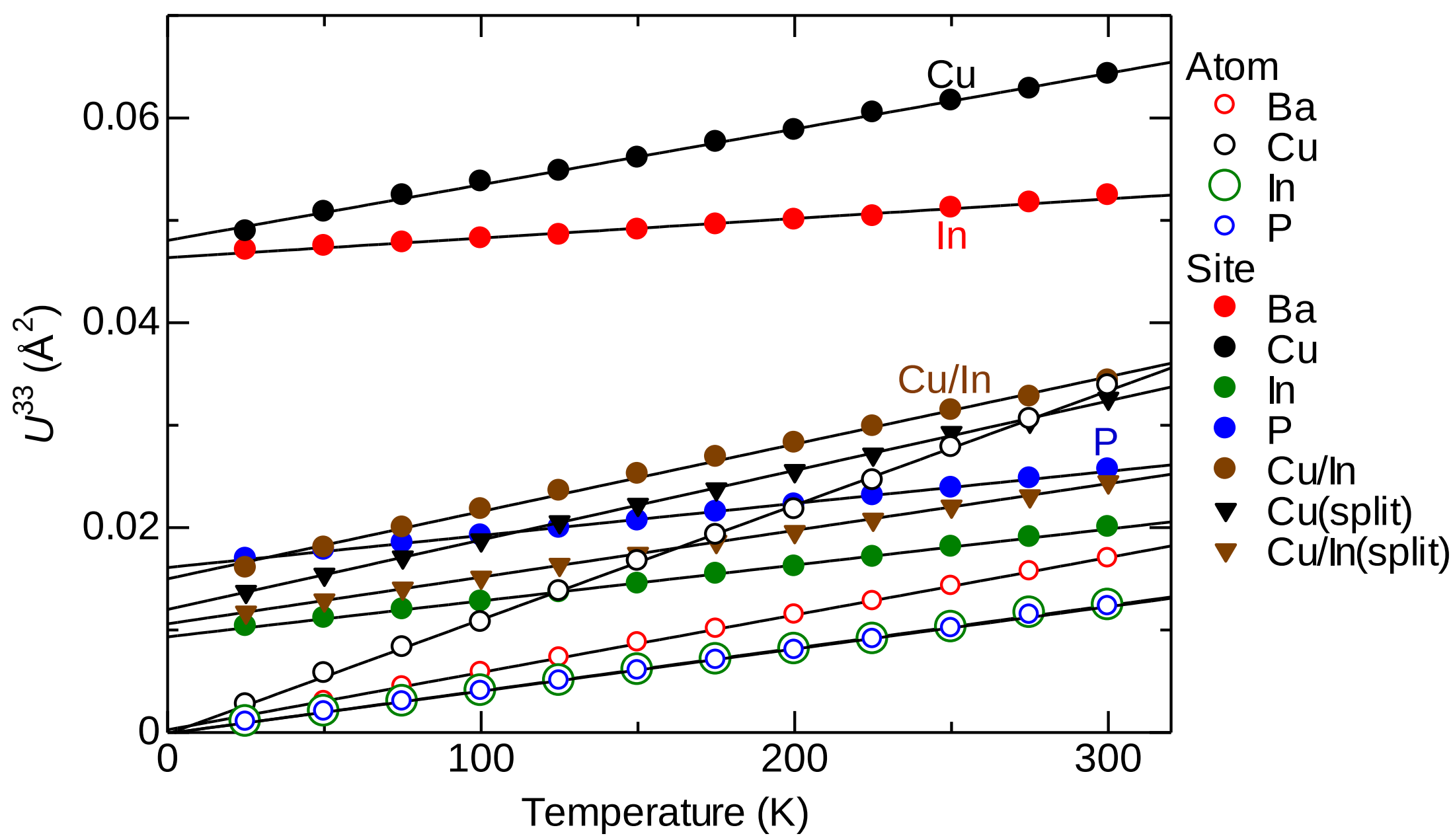

Fig. 10. Calculated $U^{33}$ per atom (empty symbols) and by site (filled symbols) in $BaCu_{1.14}In_{0.86}P_2$. The $U^{33}$ by site is the variance of the normal distribution fit as in Fig. 8. The variance from fitting of three superimposed normal distribution with same valences are shown for Cu and Cu/In sites (triangular symbols for "split" sites).